\documentclass[linenumbers]{aastex631}
\nolinenumbers
\newcommand{\alphaturb}{\alpha_{\rm turb}}
\newcommand{\alphawind}{\alpha_{\rm wind}}
\newcommand{\ffit}{f_{\rm fit}}
\newcommand{\fgas}{F_{\rm gas}}
\newcommand{\fgaso}{F_{\rm gas,0}}
\newcommand{\fice}{f_{\rm ice}}
\newcommand{\fpeb}{F_{\rm peb}}
\newcommand{\fpebo}{F_{\rm peb,0}}
\newcommand{\hgas}{H_{\rm gas}}
\newcommand{\hpeb}{H_{\rm peb}}
\newcommand{\hpml}{H_{\rm pml}}
\newcommand{\lamcrit}{\lambda_{\rm crit}}
\newcommand{\lstar}{L_\ast}
\newcommand{\mdisko}{M_{\rm disk,0}}
\newcommand{\memb}{M_{\rm emb}}
\newcommand{\membo}{M_{\rm emb,0}}
\newcommand{\miso}{M_{\rm iso}}
\newcommand{\mjup}{M_{\rm Jup}}
\newcommand{\mkb}{M_{\rm KB}}
\newcommand{\mmerc}{M_{\rm Merc}}
\newcommand{\mstar}{M_\ast}
\newcommand{\mterr}{M_{\rm terr}}
\newcommand{\mtot}{M_{\rm tot}}
\newcommand{\mtotice}{M_{\rm tot,ice}}
\newcommand{\mtotpml}{M_{\rm tot,pml}}
\newcommand{\pvs}{P_{\rm VS}}
\newcommand{\qwind}{Q_{\rm wind}}
\newcommand{\rcap}{R_{\rm cap}}
\newcommand{\remb}{R_{\rm emb}}
\newcommand{\rpml}{R_{\rm pml}}
\newcommand{\rhogas}{\rho_{\rm gas}}
\newcommand{\sigmagas}{\Sigma_{\rm gas}}
\newcommand{\sigmapeb}{\Sigma_{\rm peb}}
\newcommand{\sigmapml}{\Sigma_{\rm pml}}
\newcommand{\stefan}{\sigma_{\rm SB}}
\newcommand{\st}{{\rm St}}
\newcommand{\stcrit}{{\rm St}_{\rm crit}}
\newcommand{\stmin}{{\rm St}_{\rm min}}
\newcommand{\tauwind}{\tau_{\rm wind}}
\newcommand{\tdisk}{t_{\rm disk}}
\newcommand{\tdrag}{t_{\rm drag}}
\newcommand{\tgrow}{t_{\rm grow}}
\newcommand{\tirr}{T_{\rm irr}}
\newcommand{\tiso}{t_{\rm iso}}
\newcommand{\tmid}{T_{\rm mid}}
\newcommand{\udrift}{u_{\rm drift}}
\newcommand{\ugas}{u_{\rm gas}}
\newcommand{\upeb}{u_{\rm peb}}
\newcommand{\vfrag}{v_{\rm frag}}
\newcommand{\vice}{v_{\rm ice}}
\newcommand{\vkep}{v_{\rm kep}}
\newcommand{\vrel}{v_{\rm rel}}
\newcommand{\vroc}{v_{\rm rock}}

\shorttitle{Making the Solar System}
\shortauthors{Chambers}

\begin{document}
 
\title{Making the Solar System}

\author{John Chambers}
\affiliation{Earth and Planets Laboratory \\
Carnegie Institution for Science \\
5241 Broad Branch Road NW \\
Washington, DC 20005, USA}

%
% ABSTRACT
%
\begin{abstract}
\nolinenumbers
We model the early stages of planet formation in the Solar System, including continual planetesimal formation, and planetesimal and pebble accretion onto planetary embryos in an evolving disk driven by a disk wind. The aim is to constrain aspects of planet formation that have large uncertainties by matching key characteristics of the Solar System. The model produces a good fit to these characteristics for a narrow range of parameter space. Planetary growth beyond the ice line is dominated by pebble accretion. Planetesimal accretion is more important inside the ice line. Pebble accretion inside the ice line is slowed by higher temperatures, partial removal of inflowing pebbles by planetesimal formation and pebble accretion further out in the disk, and increased radial velocities due to gas advection. The terrestrial planets are prevented from accreting much water ice because embryos beyond the ice line reach the pebble isolation mass before the ice line enters the terrestrial-planet region. When only pebble accretion is considered, embryos typically remain near their initial mass or grow to the pebble-isolation mass. Adding planetesimal accretion allows Mars-sized objects to form inside the ice line, and allows giant-planet cores to form over a wider region beyond the ice line. In the region occupied by Mercury, pebble Stokes numbers are small. This delays the formation of embryos and stunts their growth, so that only low-mass planets can form here.

\end{abstract}
%\keywords{}

%
% INTRODUCTION
%
\section{Introduction}
The recent discovery of several thousand extrasolar planetary systems has spurred great interest in how these systems formed. While this emphasis on exoplanets is understandable, it is worth bearing in mind that we still lack a detailed model for how our own planetary system formed. This unresolved problem is made more compelling by the fact that the Solar System differs from most known exoplanet systems. The absence of super-Earth-mass planets on short-period orbits in the Solar System is particularly noteworthy given that such planets are common elsewhere.

There are good reasons to focus on the origin of our planetary system. For example, it is the only system known to host a habitable planet. We also know much more about the Solar System than other systems. We have a complete census of planetary-mass bodies out to at least 50 AU, as well as much more detailed information about the orbits of minor planet populations in our system than elsewhere. In addition, the Solar System is the only system for which we have samples that can be attributed to particular source objects. All this information provides a strong test for planet-formation theories.

The conventional model for planet formation begins with dust grains in a protoplanetary disk \citep{raymond:2014}. These grains collide and stick to form mm-to-dm sized aggregates commonly called pebbles. Further sticking is problematic due to increasing collision speeds and decreasing sticking-area-to-mass ratios \citep{blum:2008}. Instead, it is thought that instabilities in the gas disk act to collect solid particles into small regions where their collective gravity causes a collapse to form asteroid-sized bodies called planetesimals \citep{johansen:2007, johansen:2012}. 

Further growth can take place in two ways. The largest planetesimals, also known as planetary embryos, can grow by sweeping up planetesimals or pebbles. The accretion of pebbles is made more effective by drag with the disk gas, so that pebbles lose energy during an encounter with an embryo \citep{ormel:2010}. Planetesimal accretion is typically slower, especially far from the star \citep{morbidelli:2015}. For this reason, pebble accretion is thought to be the main driver of growth in the outer parts of disks \citep{lambrechts:2012}, while both mechanisms may be important in the inner disk.

Pebbles are expected to drift radially inwards over time due to gas drag \citep{weidenschilling:1977}. This drift provides a continuous supply of material for embryos to grow by pebble accretion, but it also removes pebbles that are not accreted quickly enough. Thus, pebble accretion can form massive planets by accessing material from a large fraction of the disk, but the process is relatively inefficient so a massive disk is required. When an embryo grows to a certain mass, it will perturb the local gas pressure profile sufficiently to halt the influx of pebbles \citep{lambrechts:2014}, although some smaller dust grains may continue to flow across the barrier \citep{weber:2018}. This ``pebble isolation'' stops pebble accretion onto the embryo, and also halts the supply of pebbles to embryos located closer to the star.

Embryos that reach to a different critical mass, of very roughly 10 Earth masses, begin to accrete gas at an increasing rate \citep{pollack:1996}. If the disk lasts long enough, these objects can grow into gas giant planets like Jupiter. Throughout the planet formation process, large embryos can also migrate radially due to tidal interactions with the gas disk \citep{ward:1997}.

In this paper, we examine the conditions necessary to form a planetary system that resembles the Solar System. We focus on the inner 30 AU of the solar nebula where we assume that planet formation was effective. We consider only the early stages of growth in order to avoid current uncertainties in the details of gas accretion and orbital migration. We use a simple model that includes planetesimal and planetary embryo formation, and the growth of embryos by both pebble and planetesimal accretion. Planetary growth takes place in an evolving disk, with gas and pebbles supplied at the outer edge of the planet forming region at rates that decline over time.

Planetesimals are assumed to form continually in regions where there is an influx of pebbles. We adopt the planetesimal formation model of \citet{lenz:2019}, in which planetesimals can form anywhere in the disk, at a rate proportional to the pebble flux, provided that pebbles exceed a minimum size needed to trigger instabiltilies. \citet{lenz:2020} used this model to examine the conditions needed to form enough planetesimals to build the planets and small-body populations in the Solar System. Here, we extend this kind of analysis by adding pebble accretion and considering the timescales for planetary growth.

We assume that the gas evolution is driven mainly by a disk wind \citep{suzuki:2016}. The wind efficiently removes angular momentum from the disk, causing inward gas advection, while generating weak turbulence in the gas.

The goal of this paper is to understand the conditions needed to form planetary systems that resemble our own. We do this by varying physical aspects of planet formation that are uncertain rather than using a fixed model with variable initial conditions. The current emphasis is on the Solar System due to the wealth of data we have for this system. The same approach could also be applied in future to extrasolar systems, and it is possible that some of the observed diversity of planetary systems reflects differences in the physics of formation in each case.

While the Solar System has numerous features, we concentrate on a few details that seem particularly significant. We seek systems that will form at least one gas giant in the outer disk, while the inner planets remain small and highly depleted in volatile materials, like the terrestrial planets. The innermost region, occupied by Mercury in the Solar System, should contain little solid mass, far less than needed to form a super-Earth. We want planetesimal formation to be effective in the outer Solar System since there is good evidence that a massive proto-Kuiper belt once existed here \citep{nesvorny:2018}. Finally, we want at least one embryo to reach the pebble isolation mass at an early stage in order to establish two isolated pebble and planetesimal reservoirs that could give rise to the two isotopically distinct groups of material observed in meteorites \citep{kruijer:2017}.

The rest of this paper is organized as follows. Section~2 describes the planet-formation model. Section~3 describes the requirements for Solar System analogs, and the procedure used to find sets of model parameters that satisfy these conditions. In Section~4, we look at the parameters needed to form Solar System analogs and examine the evolution in such cases. Section~5 looks at the effect of shifting the parameters away from the best fit values. Section~6 contains a discussion. The main results are summarized in Section~7.

%
% MODEL
%
\section{Model}

%
% GAS DISK
%
\subsection{Gas Disk}
We consider a protoplanetary disk that evolves due to angular momentum lost to a disk wind. As a result, gas flows inwards with a mass flux $\fgas$. The disk is heated by gravitational energy released by the inflowing gas together with reprocessed radiation from the central star. Following \citet{kondo:2022}, we assume that heat generated by the disk wind is released in a layer at a particular optical depth $\tauwind$ with respect to the disk surface.

The midplane temperature $\tmid$ at a distance $a$ from the star is
\begin{equation}
\tmid^4=\tirr^4+\frac{\qwind}{2\stefan}\left[1+\frac{3\tauwind}{4}\right]
\end{equation}
where $\stefan$ is the Stefan-Boltzmann constant, and $\qwind$ is the rate of energy release per unit disk area, given by
\begin{equation}
\qwind=\frac{\fgas\Omega^2}{4\pi}
\end{equation}
where $\Omega$ is the angular velocity.

Following \cite{chiang:1997}, the temperature $\tirr$ solely due to stellar irradiation, at a distance $a$, is
\begin{equation}
\tirr^4=\frac{2}{7}
\left(\frac{k}{G\mstar\mu m_H}\right)^{4/7}
\left(\frac{\lstar}{4\pi\stefan}\right)^{8/7}a^{-12/7}
\end{equation}
where $\mstar$ and $\lstar$ are the stellar mass and luminosity, $k$ is Boltzmann's constant, $m_H$ is the mass of a hydrogen atom, and  $\mu$ is the mean molecular weight of the gas. Following \citet{chambers:2009}, we assume that the luminosity of a solar-mass protostar at time $t$ is given by
\begin{equation}
\lstar=9.2\left(1+\frac{t}{0.1\ {\rm My}}\right)^{-2/3}L_\odot
\end{equation}

Following \citet{suzuki:2016}, the inward radial velocity of the gas is given by
\begin{equation}
\ugas=\alphawind c_s
\end{equation}
where $c_s$ is the sound speed of the gas, and $\alphawind$ is a measure of the strength of the disk wind. The gas surface density is
\begin{equation}
\sigmagas=\frac{\fgas}{2\pi a\ugas}
\end{equation}

Following \citet{hartmann:2018}, we assume that the inward gas flux varies inversely with time $t$ following an infall phase lasting for $\tdisk$. The gas flux is given by
\begin{equation}
\fgas=\fgaso\left[1+\frac{t}{\tdisk}\right]^{-1}
\end{equation}

%
% PEBBLES
%
\subsection{Pebbles}
We assume that pebbles are injected into the planet-forming region of the disk from larger distances with a mass flux
\begin{equation}
\fpebo=Z\fgas
\end{equation}
where $Z$ is the dust-to-gas ratio, and we assume that the gas and pebble fluxes decline at the same rate following \citet{lambrechts:2019}.

The pebble flux $\fpeb$ at smaller distances is reduced with respect to $\fpebo$ as a result of conversion of pebbles into planetesimals and pebble accretion onto embryos, which are calculated at each location in the disk (see below).

Following \citet{dullemond:2018}, pebbles move radially inwards at a speed $\upeb$ that depends on the inward advection of the gas $\ugas$, and the drift rate $\udrift$ with respect to the gas
\begin{equation}
\upeb=\udrift+\frac{\ugas}{(1+\st^2)}
\label{eq-upeb}
\end{equation}
where $\st$ is the pebble Stokes number, and
\begin{equation}
\udrift=\frac{\st}{(1+\st^2)}\frac{c_s^2}{\vkep}
\frac{d\ln P}{d\ln a}
\end{equation}
where $P$ is the gas pressure, and $\vkep=a\Omega$ is the Keplerian velocity.

Thus, the pebble surface density at distance $a$ is given by
\begin{equation}
\sigmapeb=\frac{\fpeb}{2\pi a\upeb}
\end{equation}

We assume that pebbles grow quickly by mutual collisions until they reach a maximum size at which collisions begin to cause erosion. Collisional velocities are assumed to be set mostly by turbulence in the gas, such that the pebble Stokes number is given by
\begin{equation}
\st=\frac{\vfrag^2}{3\alphaturb c_s^2}
\end{equation}
\citep{ormel:2007}, where $\alphaturb$ is a measure of the turbulence strength, and $\vfrag$ is the collision velocity at which collisions become erosional. We allow for different material strengths inside and outside the ice line, assumed to lie at the distance where $\tmid=160$ K, such that
\begin{eqnarray}
\vfrag&=&\vroc \hspace{20mm} T>160\ K \nonumber \\
&=&\vice \hspace{22mm} T<160\ K 
\end{eqnarray}
We use a fixed value of $\vroc=100$ cm/s, which is a typical experimental result for silicates \citep{guttler:2010}. The ice fragmentation speed is a model parameter.

In addition, we assume that the pebble flux decreases by a factor of 2 moving inwards across the ice line due to evaporation of the icy component.

Following \citet{bitsch:2018}, an embryo with a mass exceeding the pebble isolation mass $\miso$ will halt the inward flux of pebbles. As a result, pebble accretion and planetesimal formation will be halted at this distance and all distances closer to the star, where
\begin{equation}
\miso=\left(25+\frac{\lamcrit}{0.00476}\right)\ffit M_\oplus
\end{equation}
where
\begin{equation}
\lamcrit=\frac{\alphaturb}{2\st}
\end{equation}
and
\begin{equation}
\ffit=\left(\frac{\hgas}{0.05a}\right)^3
\left[0.34\left(\frac{-3}{\log_{10}\alphaturb}\right)^4+0.66\right]
\left[\frac{7}{12}-\frac{1}{6}\frac{d\ln P}{d\ln a}\right]
\end{equation}

%
% PLANETESIMAL AND EMBRYO FORMATION
%
\subsection{Planetesimal and Embryo Formation}
We use the model of \citet{lenz:2019} for the formation of planetesimals and planetary embryos. Planetesimals are assumed to form continually from pebbles in most regions of the disk such that the planetesimal surface density $\sigmapml$ increases at a rate
\begin{eqnarray}
\frac{d\sigmapml}{dt}&=&\left(\frac{\epsilon\udrift}{5\hgas}\right)\sigmapeb
\hspace{20mm} \st>\stmin \nonumber \\
&=&0 \hspace{41mm} \st<\stmin
\end{eqnarray}
where $\hgas$ is the gas scale height, and $\epsilon$ is an efficiency factor for the conversion of pebbles into planetesimals. Following \citet{lenz:2019}, we assume that planetesimal formation only occurs for pebbles with a Stokes number above a critical value $\stmin$. Both $\epsilon$ and $\stmin$ are parameters in our simulations.

Numerical simulations suggest that planetesimals form with a range of sizes, typically $\sim100$ km, with a decreasing probability at larger sizes \citep{johansen:2015, simon:2016, li:2019}. We model this by assuming that most planetesimals form with a radius $\rpml=50$ km. However, when planetesimals are first able to form at a particular distance, we assume that a small number of larger planetary embryos form at this location. The embryos  have initial mass $\membo$.

We ignore the formation of new embryos in a particular region after the first ones have formed. Instead, we assume that existing embryos grow by accreting pebbles and planetesimals, and embryos occasionally merge such that their mean orbital spacing is maintained at $b$ Hill radii, where $b\simeq10$ \citep{kokubo:1998}.

%
% PEBBLE ACCRETION
%
\subsection{Pebble Accretion}
Following \citet{ormel:2010}, a planetary embryo of mass $\memb$ accretes pebbles at a rate
\begin{equation}
\frac{d\memb}{dt}=\sigmapeb\vrel\times\max\left[
\frac{\pi\rcap^2}{2\hpeb},2\rcap\right]
\end{equation}
where $\hpeb$ is the scale height of the pebbles, given by \citet{youdin:2007}
\begin{equation}
\hpeb=\hgas\left(\frac{\alphaturb}{\alphaturb+\st}\right)^{1/2}
\end{equation}
and $\rcap$ is the capture radius for pebble accretion, given by
\begin{equation}
\rcap=r_H\times\min\left[
\left(\frac{12r_H\st}{a\eta}\right)^{1/2},2\st^{1/3}\right]
\exp\left[-\left(\frac{\st}{\stcrit}\right)^{0.65}\right]
\label{eq-rcap}
\end{equation}
where $r_H$ is the embryo's Hill radius, and 
\begin{equation}
\eta=\frac{1}{2}\left(\frac{c_s}{\vkep}\right)^2\frac{d\ln P}{d\ln a}
\end{equation}

The exponential term in Eqn.~\ref{eq-rcap} is an empirical factor that accounts for the reduced efficiency of pebble accretion for large pebbles, where
\begin{equation}
\stcrit=\min\left[12\left(\frac{r_H}{a\eta}\right)^3,\frac{2}{3}
\right]
\label{eq-stcrit}
\end{equation}

In addition, $\vrel$ is the pebble-embryo relative velocity just before an encounter, given by
\begin{equation}
\vrel=\vkep\times\max\left[\eta,\frac{3r_H\st^{1/3}}{a}\right]
\end{equation}

%
% PLANETESIMAL ACCRETION
%
\subsection{Planetesimal Accretion}
Following \citet{chambers:2014}, an embryo accretes planetesimals at a rate
\begin{equation}
\frac{d\memb}{dt}=\frac{\pi\remb^2\vrel\sigmapml}{2\hpml}
\left[1+\frac{2G\memb}{\remb\vrel^2}\right]
\end{equation}
where $\remb$ is the radius of the embryo, $\sigmapml$ is the surface density of planetesimals, $\vrel$ is the relative velocity shortly before an encounter, given by
\begin{equation}
\vrel=\vkep\times\max\left[e,\frac{r_H}{a}\right]
\end{equation}
where $e$ is the mean planetesimal eccentricity, and $\hpml$ is the  planetesimal scale height, given by
\begin{equation}
\hpml=ai\simeq\frac{ae}{2}
\end{equation}
where $i$ is the mean planetesimal inclination, and the last approximation is appropriate for planetesimal in the dispersion dominated regime.

We calculate the evolution of $e$ taking into account viscous stirring from the embryos and damping due to gas drag. The viscous stirring rate is
\begin{equation}
\frac{de^2}{dt}=\frac{\memb\pvs}{3\mstar bP}
\end{equation}
where $P$ is the orbital period, $b$ is the mean orbital spacing of embryos in Hill radii, and
\begin{equation}
\pvs=73\left[1+0.11\frac{a^2e^2}{r_H^2}\right]^{-1}
\end{equation}

The gas drag damping rate is
\begin{equation}
\frac{de^2}{dt}=-\frac{2e^2}{\tdrag}\left[1+\left(\frac{e}{\eta}\right)^2\right]
\end{equation}
where
\begin{equation}
\tdrag=\frac{6\rho\rpml}{\rhogas\eta\vkep}
\end{equation}
where $\rho$ is the bulk density, and $\rhogas$ is the gas density.

We neglect radial scattering and radial drift of planetesimals.

%
% MODEL PARAMETERS AND SIMULATION DETAILS
%
\subsection{Model Parameters and Simulation Details}
We model the formation and evolution of systems like the Solar System by considering the formation of planetesimals, and the accretion of pebbles and planetesimals onto planetary embryos in an evolving protoplanetary disk driven by a disk wind. We focus on the inner 30 AU of the disk since this is where the Sun's planets are located. We are interested primarily on processes that occur while the disk is still present, so simulations are terminated after 3 My.

The disk is divided into 200 logarithmically spaced radial zones, spanning 0.1 to 30 AU. Gas and pebbles are assumed to flow into this region from the outer edge, being supplied by more distant parts of the disk that we do not model here. The time-integrated gas and pebble fluxes are $0.086M_\odot$ and $290M_\oplus$, respectively, over the entire simulation. In each zone, we track the inward flux of pebbles, the conversion of some pebbles into planetesimals, and the accretion of both populations by planetary embryos.

At each step in the evolution, we calculate the number of embryos (possibly $<1$) per radial zone. Embryos in the same zone are assumed to have identical masses and growth rates rather than following the growth of each embryo separately. We note that orbital migration and gas accretion onto embryos are not included for reasons described in the Introduction. The embryos are assumed to maintain a constant spacing in Hill radii due to mergers and scattering although embryo-embryo interactions are not calculated explicitly.

In this study, we concentrate on 6 parameters with values that are especially uncertain. These are (i) the strength of disk evolution driven by the disk wind, quantified by $\alphawind$; (ii) the strength of turbulence in the gas, quantified by $\alphaturb$; (iii) the optical depth of the disk layer that is heated by the disk wind, measured from the surface, denoted by $\tauwind$; (iv) the fragmentation velocity of icy pebbles, denoted by $\vice$; (v) the conversion efficiency from pebbles to planetesimals, quantified by $\epsilon$; and (vi) the minimum Stokes number of pebbles that can be converted into planetesimals, denoted by $\stmin$.

These 6 parameters, together with the range of values that we explore, are listed in Table~1. The remaining model parameters, which we keep fixed, are listed in Table~2.

\begin{deluxetable*}{lccc}
\tablenum{1}
\tablecaption{Variable model parameters and permitted values}
\tablewidth{0pt}
\tablehead{
\colhead{Parameter} & Symbol & \colhead{Value} &
\colhead{Best Fit}
}
\decimalcolnumbers
\startdata
Turbulence strength & $\alphaturb$ & $10^{-5}$--$10^{-3}$ &
$6.75\pm0.37\times 10^{-5}$ \\
Disk wind strength & $\alphawind$ & $10^{-3}$--$10^{-2}$ &
 $1.03\pm0.04\times10^{-3}$ \\
Optical depth of heated layer & $\tauwind$ & 1--30 &
$22.7\pm1.8$ \\
Planetesimal formation efficiency & $\epsilon$ & $10^{-3}$--1 &
$0.0407\pm0.0003$ \\
Minimum $\st$ for planetesimals to form & $\stmin$ & $10^{-3}$--0.1 
& $3.53\pm0.29\times 10^{-3}$ \\
Ice fragment speed & $\vice$ & 20--500 cm/s &
$186\pm12$ cm/s \\
\enddata
\end{deluxetable*}

\begin{deluxetable*}{lcc}
\tablenum{2}
\tablecaption{Fixed model parameters and their values}
\tablewidth{0pt}
\tablehead{
\colhead{Parameter} & Symbol & \colhead{Value}
}
\decimalcolnumbers
\startdata
Stellar mass & $\mstar$ & $M_\odot$ \\
Initial gas flux & $\fgaso$ & $10^{-7}M_\odot$/y \\
Gas infall time & $\tdisk$ & 0.4 My \\
Metallicity & $Z$ & 0.01 \\
Bulk density & $\rho$ & 2 g/cm$^2$ \\ 
Gas mean molecular weight & $\mu$ & 2.3 \\
Rock fragment speed & $\vroc$ & 100 cm/s \\
Planetesimal radius & $\rpml$ & 50 km \\
Initial embryo mass &$\membo$ & $10^{-5}M_\oplus$ \\
Mean embryo separation & $b$ & 10 $r_H$ \\
\enddata
\end{deluxetable*}

%
% SOLAR SYSTEM CONSTRAINTS
%
\section{Solar System Constraints}
In this section, we devise a short list of constraints that a planetary system should satisfy in order to be considered ``Solar-System like''. Additional constraints could certainly be added to this list, but the ones given below were chosen to be at least somewhat model independent. We also describe the procedure used to find sets of parameters that match these constraints.

%
% MASS OF THE TERRESTRIAL PLANETS
%
\subsection{Mass of the Terrestrial Planets} 
The combined mass of the inner planets of the Solar System is approximately 2 Earth masses, and their orbits span roughly 0.4--1.5 AU. It is unlikely that the inner planets had finished growing at the point when the solar nebula dispersed, but most of the material that would end up in these planets was probably present in this region at that time. 

Thus, we require that the total mass $\mterr$ of embryos and planetesimals interior to 1.5 AU at the end of the simulation is about 2 Earth masses.

%
% MASS IN THE REGION OCCUPIED BY MERCURY
%
\subsection{Mass in the Region Occupied by Mercury} 
Most of the mass contained in the terrestrial planets exists in Venus and Earth, which occupy a relatively narrow region in terms of distance from the Sun. It is plausible that the region beyond Earth's orbit was depleted by processes associated with Jupiter \citep{walsh:2011, clement:2018}, which we do not attempt to model here. The relative lack of material interior to Venus requires a different explanation. Here, we assume that planetesimal formation and/or mass accretion onto planetary embryos was inefficient in this region, but not entirely prevented. 

Specifically, we require that the total mass $\mmerc$ of embryos and planetesimals interior to 0.5 AU is 0.025--0.1 Earth masses, which is roughly within a factor of 2 of Mercury's mass.

%
% WATER FRACTION OF THE TERRESTRIAL PLANETS
%
\subsection{Water Fraction of the Terrestrial Planets}
The terrestrial planets  contain very little water compared to the roughly solar composition of the solar nebula. The combined mass of water in Earth's oceans and interior is almost certainly $<1$ \% of the planet's mass, while the other inner planets are drier still \citep{marty:2012,halliday:2013}. While the origin of this depletion remains a matter of debate, here we assume it is the result of materials accreted by the terrestrial planets. We neglect the possibilty that water delivered by pebbles is  not accreted by a planetary embryo due to evaporation in the envelope or refluxing back to the nebula \citep{chambers:2017,ali-dib:2020,johansen:2021}.

Thus, we require that the water mass fraction $\fice$ of the embryos and planetesimals interior to 1.5 AU at the end of the simulation is $<1$ \%.

%
% MASS OF JUPITER'S CORE
%
\subsection{Mass of Jupiter's Core}
In the core accretion model, Jupiter's growth begins with the formation of a solid core that grows too massive to sustain a hydrostatic atmosphere, thereby initiating runaway gas accretion. Simulations of core accretion typically find that runaway gas accretion requires a core of roughly 10 Earth masses \citep{pollack:1996,ikoma:2000}. We do not consider the possibility that ongoing pebble accretion may delay the onset of gas accretion.

Jupiter's core must form before the solar nebula dispersed, and early enough to allow time for gas accretion to occur. Here, we simply require that a 10-Earth-mass core forms at Jupiter's current location, but note that the pebble barrier constraint described below typically requires that this core forms early in the nebula's lifetime. 

We require embryos at 5 AU to have a mass $\mjup$ of at least 10 Earth masses at the end of the simulation.

%
% MASS OF THE PROTO-KUIPER BELT
%
\subsection{Mass of Proto-Kuiper Belt} 
The Nice model for the early evolution of the Solar System indicates that the orbits of the giant planets migrated radially after the planets formed due to interactions with a disk of planetesimals at larger distances \citep{tsiganis:2005}. These planetesimals failed to form an additional planet, and the great majority were ultimately ejected from the Solar System, with the remnants forming the modern Kuiper belt. The Nice model suggests that roughly 15--20 Earth masses of planetesimals existed beyond Neptune's original orbit, occupying a region of roughly 20--30 AU from the Sun \citep{nesvorny:2018}.

Thus we require that the mass $\mkb$ of planetesimals between 20 and 30 AU is 15--20 Earth masses at the end of a simulation.

%
% AN EARLY BARRIER TO PEBBLE DRIFT
%
\subsection{An Early Barrier to Pebble Drift}
Meteorites can be divided into two distinct groups on the basis of the isotopic ratios of a number of elements \citep{kleine:2020}. This dichotomy spans meteorites from differentiated and undifferentiated parent bodies. The estimated ages for the two groups of parent bodies overlap, which suggests the meteorites sample planetesimals that formed or were modified in two isolated regions of the solar nebula. The origin of the dichotomy is uncertain, but meteorite data imply it was established by roughly 0.5 My after the first solids formed \citep{kleine:2020}. Here, we adopt a model in which the dichotomy is established and maintained when an embryo becomes massive enough to block the inward flow of pebbles \citep{kruijer:2017}.

Thus we require that the time $\tiso$ needed for an embryo to reach the pebble isolation mass somewhere in the disk is no more than 0.5 My.

%
% QUANTIFYING THE FIT
%
\subsection{Quantifying the Fit}
Based on the constraints described above, we assign a score $S$ to the output of a simulation as follows
\begin{equation}
S=S_1\times S_2\times S_3\times S_4\times S_5\times S_6
\label{eq-score}
\end{equation}
where $S_1$ is a score based on the combined mass of the terrestrial planets:
\begin{eqnarray}
S_1&=&\min\left[x_1,\frac{1}{x_1}\right] \nonumber \\
x_1&=&\frac{\mtot(a<1.5\ {\rm AU})}{2\ M_\oplus}
\end{eqnarray}
$S_2$ is a score based on the mass in the region occupied by Mercury:
\begin{eqnarray}
S_2&=&\min[1,4x_2] \hspace{20mm} x_2<1 \nonumber \\
&=&\frac{1}{x_2} \hspace{32mm} x_2\ge 1 \nonumber \\
x_2&=&\frac{\mtot(a<0.5\ {\rm AU})}{0.1\ M_\oplus}
\end{eqnarray}
$S_3$ is a score based on the water mass fraction of the terrestrial planets
\begin{eqnarray}
S_3&=&\min\left[1,\frac{1}{x_3}\right] \nonumber \\
x_3&=&\frac{\mtotice(a<1.5\ {\rm AU})}{0.01\mtot(a<1.5\ {\rm AU})}
\end{eqnarray}
$S_4$ is a score based on the mass of Jupiter's core:
\begin{eqnarray}
S_4&=&\min[1,x_4] \nonumber \\
x_4&=&\frac{\memb(a=5\ {\rm AU})}{10\ M_\oplus}
\end{eqnarray}
$S_5$ is a score based on the mass of the primordial Kuiper belt:
\begin{eqnarray}
S_5&=&\min\left[1,\frac{4x_5}{3}\right]
\hspace{20mm} x_5<1 \nonumber \\
&=&\frac{1}{x_5} \hspace{36mm} x_5\ge 1 \nonumber \\
x_5&=&\frac{\mtotpml(20<a<30\ {\rm AU})}{20\ M_\oplus}
\end{eqnarray}
and $S_6$ is a score based on the time $\tiso$ at which the first embryo reaches the pebble isolation mass:
\begin{eqnarray}
S_6&=&\min\left[1,\frac{1}{x_6}\right] \nonumber \\
x_6&=&\frac{\tiso}{0.5\ {\rm My}}
\end{eqnarray}

\begin{figure}
\plotone{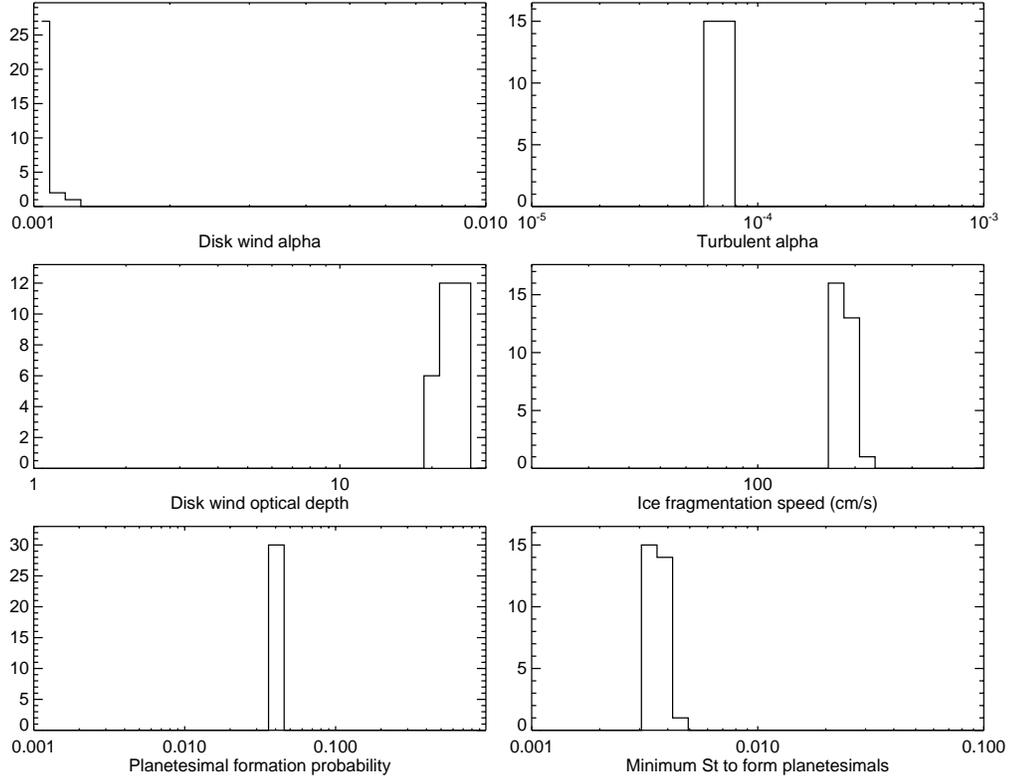}
\caption{Distribution of parameter values for simulations that approximately satisfy the Solar System constraints described in Section~3. All simulations have a score $S>0.98$.}
\end{figure}

%
% OPTIMIZATION STRATEGY
%
\subsection{Optimization Strategy}
Using the planet formation model described in Section 2, we search for values of the 6 model parameters listed in Table~1 that yield systems that satisfy the Solar System constraints listed above. 

Following \citet{chambers:2018}, we use a particle swarm optimization (PSO) scheme in which a swarm of particles explores the 6-dimensional phase space of parameter values. Each particle has an instantaneous position ${\bf x}$ and velocity ${\bf v}$ in phase space. We also keep track of the best-fit set of parameters ${\bf p}$ found by each particle, and the global best fit ${\bf g}$ found by any of the particles.

The PSO scheme undergoes a series of iterations. At each iteration, and for each particle $i$, we run a planet-formation simulation using the set of parameters defined by ${\bf x}_i$, and calculate the score for the output of the simulation using Eqn.~\ref{eq-score}. If the fit is better than the existing fit for the particle, or for the swarm as a whole, then the stored fit is updated. The particle location is then updated using
\begin{eqnarray}
{\bf v}_i&=&C_1{\bf v}_i+C_2z_1({\bf p}_i-{\bf x}_i)
+C_3z_1({\bf g}-{\bf x}_i)
\nonumber \\
{\bf x}_i&=&{\bf x}_i+{\bf v}_i
\end{eqnarray}
where $z_1$ and $z_2$ are uniform random numbers between 0 and 1. If there is no improvement in the fit after 20 iterations, we nudge each particle a small distance in a random direction in phase space and continue. 

We use 20 swarm particles. Following \citet{chambers:2018}, we choose $C_1=0.83$ and $C_2=C_3=1.65$. Typically, a few thousand iterations are needed to find an excellent fit (score $S>0.98$) to the Solar System constraints. We performed 31 optimization runs, starting with different initial random particle positions ${\bf x}$. In all but one case, a successful fit was found.

%
% BEST-FIT CASES
%
\section{Best Fit Cases}
%
% PARAMETER DISTRIBUTIONS
%
\subsection{Parameter Distributions}
Figure 1 shows the distribution of parameter values for 30 fits found by the particle-swarm optimization (PSO) scheme. These simulations produced excellent matches to the Solar System constraints described in the previous section. In all cases, they scored at least 0.98, where 1 is the maximum possible score. The means and standard deviations for the best fit parameters from the 30 simulations are given in Table~1.

For each of the 6 parameters, the best-fit cases cluster around a particular value with a range much narrower than the range of values that were explored. However, some spread is observed. Thus, simulations that match the Solar System constraints occupy a small but finite region of phase space.

\begin{figure}
\plotone{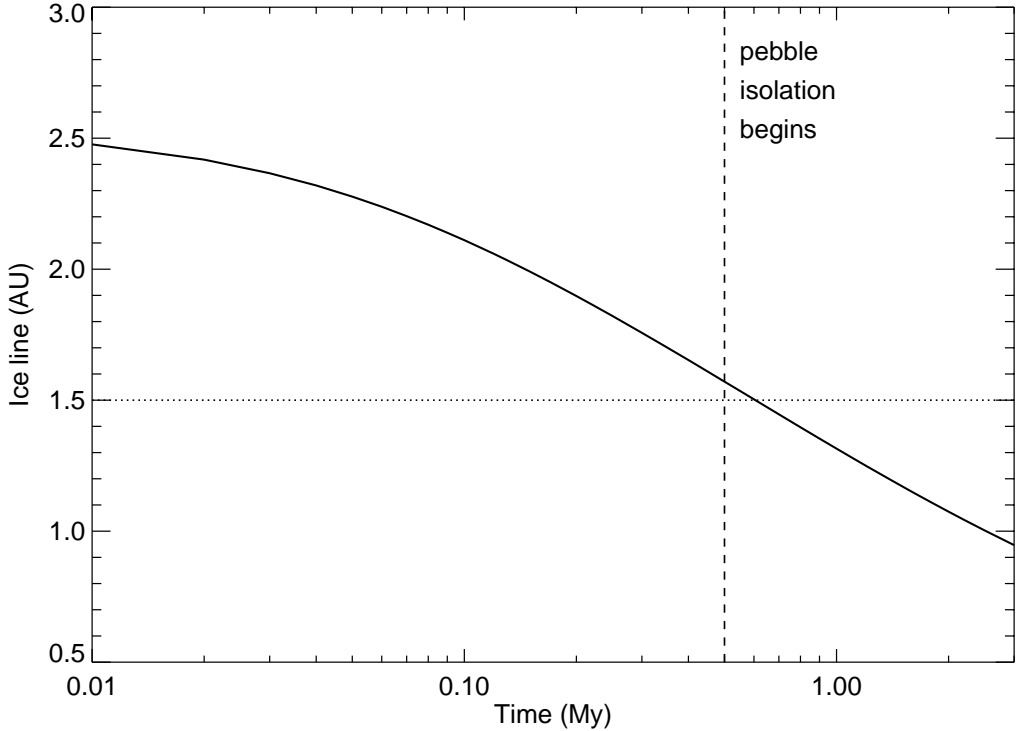}
\caption{Evolution of the water ice line in a typical best fit simulation. The vertical dashed line shows the time at which the first embryo becomes massive enough to stop the inward flux of pebbles. This occurs shortly before icy pebbles can exist in the terrestrial-planet region ($a<1.5$ AU).}
\end{figure}

The distributions for 5 of the parameters lie away from the edges of the permitted region of phase space, which implies that these boundaries did not bias the results. (The values of the optical depth parameter $\tauwind$ lie close to the highest permitted value, which is 30, but do not exceed 25.) The disk-wind parameter $\alphawind$ is an exception, with the best-fit cases clustered against $10^{-3}$ which is the lowest value considered here. 

This suggests there is an additional region of phase space, at smaller values of $\alphawind$, that satisfies the Solar System constraints . However, such values may correspond to unrealistically massive disks.  When $\alphawind=10^{-3}$, the initial disk mass interior to 30 AU is $0.026M_\odot$. Here, we assume that this region is continually supplied by material flowing inwards from more distant parts of the disk. Thus, the majority of the mass should lie outside 30 AU initially. The total disk mass is likely to be several times higher than $0.026M_\odot$, which puts the disk at the high end for observed protoplanetary disk masses \citep{tychoniec:2020}. The disk surface density is proportional to $1/\alphawind$, so values of $\alphawind<10^{-3}$ would correspond to even more massive disks. For this reason, we do not explore smaller values of $\alphawind$.

Some of the parameter values seen in Figure~1 can be readily understood. For example, the temperature structure of the inner disk is determined by $\tauwind$. If $\tauwind$ is too small, we would expect the water ice fraction of the terrestrial planets to increase above the 1\% constraint we imposed in Section~3. More specifically, we require that the ice line remains beyond the outer boundary of the terrestrial planet region (1.5 AU) until a more distant embryo becomes massive enough to stop the influx of pebbles. In the best-fit case, the first embryo reaches this pebble isolation mass at about 0.5 My. Figure 2 shows the evolution of the ice line in this case, and we see that it does indeed cross 1.5 AU at roughly 0.5 My.

The preferred value of the planetesimal formation probability $\epsilon$ also has a simple explanation, at least to first order. Provided that the pebble Stokes number exceeds $\stmin$, the mass of planetesimals that form in the Kuiper belt depends mainly on $\epsilon$. The requirement that 15--20 Earth masses of planetesimals form in the Kuiper belt constrains $\epsilon$ to a fairly narrow range of values. 

\begin{figure}
\plotone{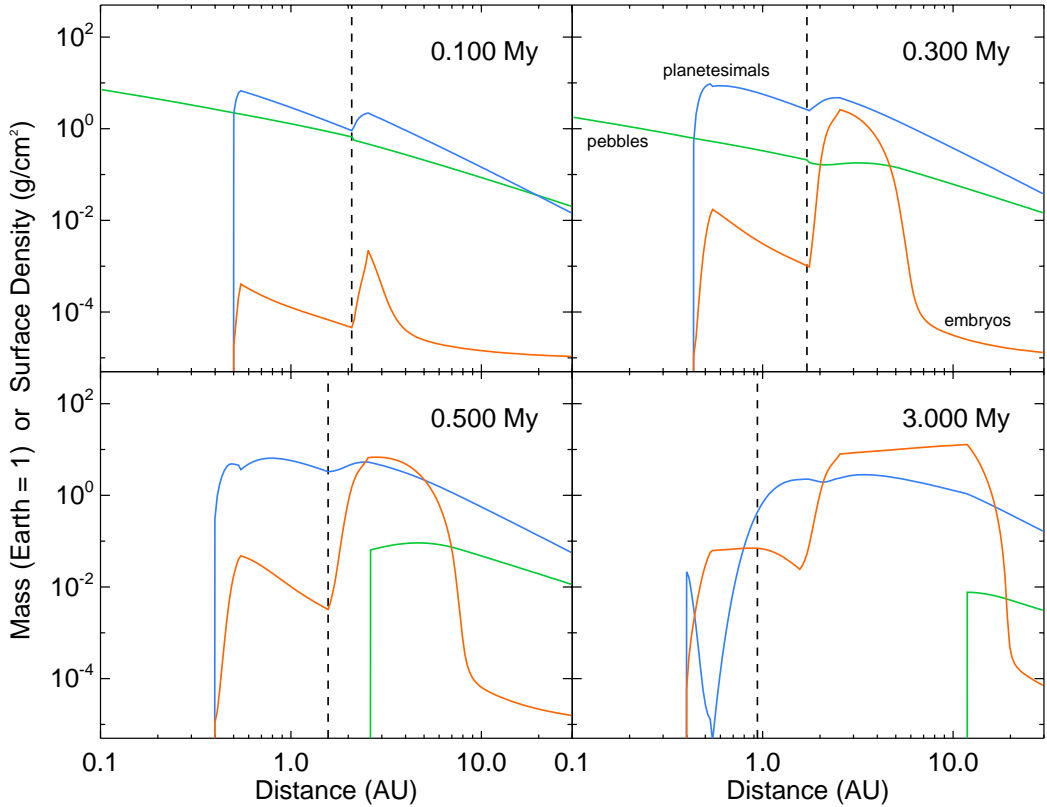}
\caption{Evolution in a typical best-fit case with $\alphaturb=7.14\times 10^{-5}$, $\alphawind=1.00\times 10^{-3}$, $\tauwind=24.9$, $\epsilon=0.0403$, $\stmin=3.15\times 10^{-3}$ and $\vice=201$ cm/s. The green and blue curves show the pebble and planetesimal surface densities versus distance at four times. The red curves show embryo mass as a function of distance. The dashed line shows the location of the water ice line. Note that planetary embryos and planetesimals fail to form inside 0.4 AU.}
\end{figure}

We can get additional insights into the best-fit parameter values by looking at the evolution in detail, as we do below, and by seeing the effect of varying the parameters, which is explored in the Section 5.

%
% EVOLUTION IN BEST-FIT CASE
%
\subsection{Evolution in Best-Fit Case}
Figure~3 shows the state of a typical successful simulation at 4 epochs in time. The green and blue curves in the figure show the pebble and planetesimal surface densities, respectively, as a function of distance. The red curves show the embryo masses at each distance.

Early in the simulation, planetesimals and embryos do not form within about 0.5 AU of the Sun. In this region, the pebble Stokes number $\st$ is below the threshhold $\stmin$ for planetesimals to form in the model used here \citep{lenz:2019}. The Stokes number is set by turbulent fragmentation, with $\st\propto 1/\tmid$. As a result, $\st$ increases with radial distance (see Figure~4), so planetesimals and embryos can form beyond a certain distance in the disk where $\st=\stmin$. The Stokes number also increases over time as the disk cools, so the inner edge of the planetesimal- and embryo-forming region moves inwards. The location and evolution of this inner edge are the main factors that control the amount of solid material in the region now occupied by Mercury.

%
% PEBBLE SURFACE DENSITY
%
\subsection{Pebble Surface Density}
For almost the first 0.5 My, pebbles are present throughout the disk, with a surface density decreasing with distance such that $\sigmapeb\propto\fpeb/(a\upeb)$. The pebble flux $\fpeb$ declines over time and so does $\sigmapeb$. There is also a small jump at the ice line caused by ice evaporation and the change in the fragmentation velocity from $\vroc$ to $\vice$.

The pebble radial velocity $\upeb$ has two components; (i) the drift of pebbles with respect to the gas $\udrift\propto a^{1/2}$, and (ii) the advection of the gas $\ugas\propto\tmid^{1/2}$, where the scalings are appropriate for the model used here. These components are shown in Figure~4. The drift component dominates outside the ice line due to the higher fragmentation strength of ice than rock in the best-fit case, which corresponds to higher $\st$ and faster drift. Inside the ice line, however, the radial motion of the gas is typically more important. The difference explains the change in slope of $\sigmapeb$ at the ice line seen in the first panel in Figure~3.

As the simulation progresses, the pebble surface density profile flattens somewhat. This is because pebble accretion becomes more efficient as embryos grow larger, so the flux of pebbles reaching the inner disk is reduced due to the loss of pebbles accreted at larger distances.

At about 0.5 My, embryos located near 2.5 AU reach the pebble isolation mass. The flux of pebbles to the inner disk is halted at this point, which can be seen in the last two panels of Figure~3. Over time, embryos at progressively larger distances also reach the pebble isolation mass (see Figure~5), so that an increasingly large fraction of the disk ceases to receive pebbles.

\begin{figure}
\plotone{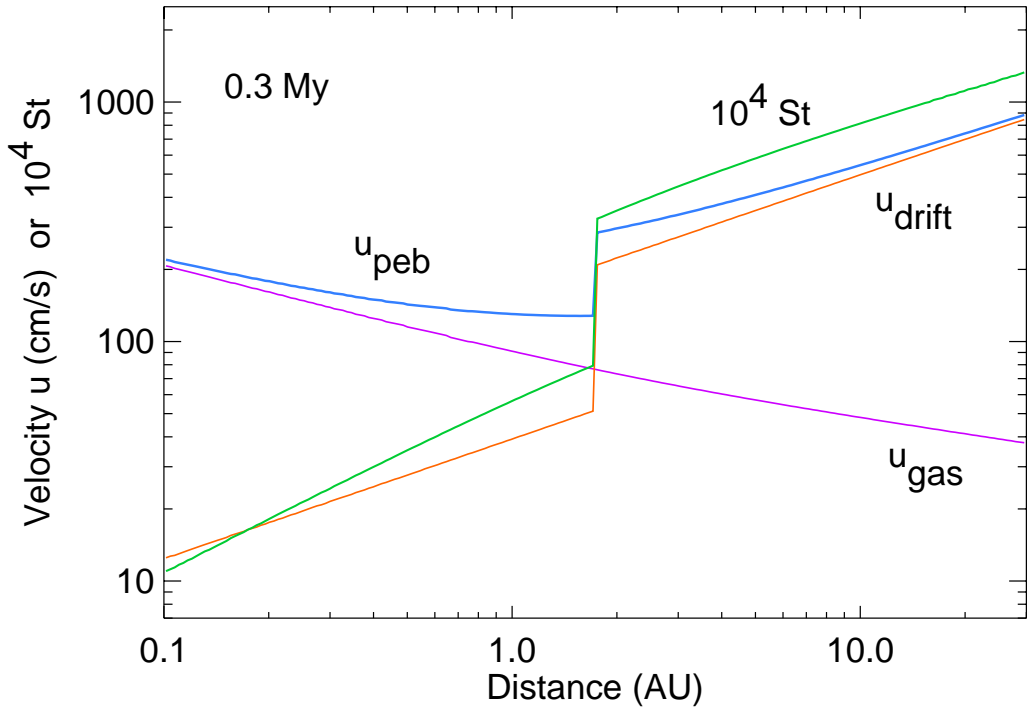}
\caption{Blue curve: total radial velocity of the pebbles at 0.3 My in the simulation shown in Figure~3. Red and purple curves: the contributions to this velocity due to pressure-driven radial drift and the motion of the gas, respectively. Green curve: the pebble Stokes number multiplied by $10^4$.}
\end{figure}

\begin{figure}
\plotone{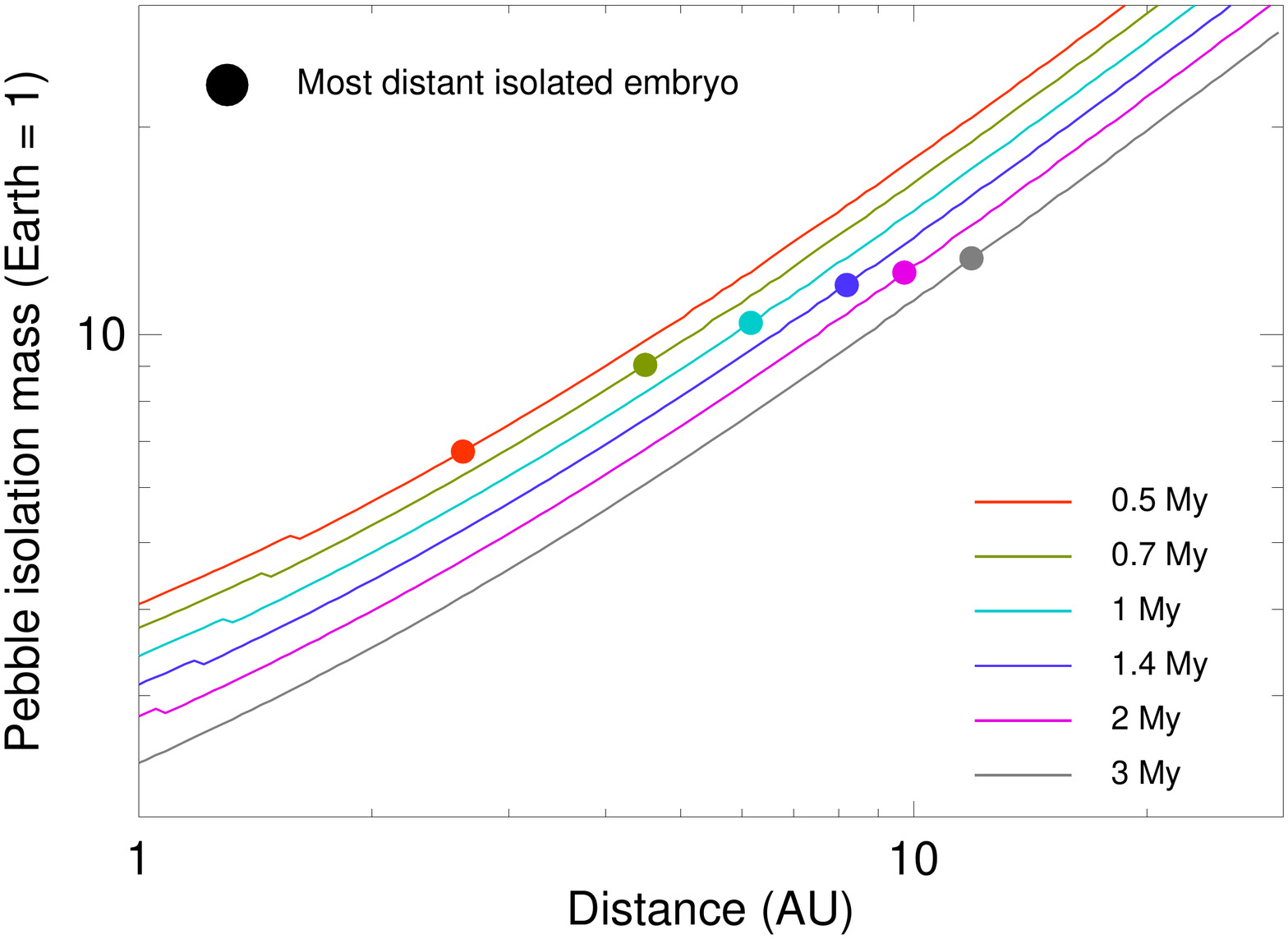}
\caption{Pebble isolation mass versus distance at 6 times for the simulation shown in Figure 3. The circular symbols show the location of the outermost embryo that has reached the pebble isolation mass at each time.}
\end{figure}

%
% PLANETESIMAL SURFACE DENSITY
%
\subsection{Planetesimal Surface Density}
The blue curves in Figure 3 show the planetesimal surface density $\sigmapml$. Planetesimals form rapidly, especially in the inner disk, except for the region where $\st<\stmin$. By 0.1 My, the surface density $\sigmapml$ exceeds the pebble surface density across most of the planet-forming region.

Planetesimals form at a rate $\propto\sigmapeb/(a\tmid^{1/2})$. The denominator in this expression increases with radial distance, which means the surface density profile for the planetesimals is steeper than that for the pebbles.

At early times, planetesimals form at a faster rate than they are accreted by embryos throughout the disk. Planetesimal accretion rates increase over time since they depend on the cumulative mass of planetesimals that have formed in a region. Conversely, planetesimal formation rates decline as the pebble flux decreases. The two rates become equal at about 0.4 My at 1 AU (see Figure 6), after which the planetesimal surface density declines. Shortly after this, the flux of pebbles into the inner disk ceases, planetesimal formation stops, and $\sigmapml$ declines more rapidly.

In the outer regions of the disk, planetesimal formation is less rapid than in the inner disk, but planetesimal accretion is even more inefficient. Few planetesimals are removed, and the surface density increases steadily over time. By 3 My, $\sigmapml$ at 25 AU is only a few times smaller than at 1 AU, as shown in Figure~6. 

The prolonged nature of planetesimal formation in the outer disk explains why a massive proto-Kuiper belt is able to form despite the fact that planetesimal formation is inefficient at large distances in the model of \citet{lenz:2019}. Conversely, planetesimals form rapidly in the inner disk, but formation stops at an early stage due to pebble isolation, and planetesimals are also swept up quite rapidly by embryos in this region.

\begin{figure}
\plotone{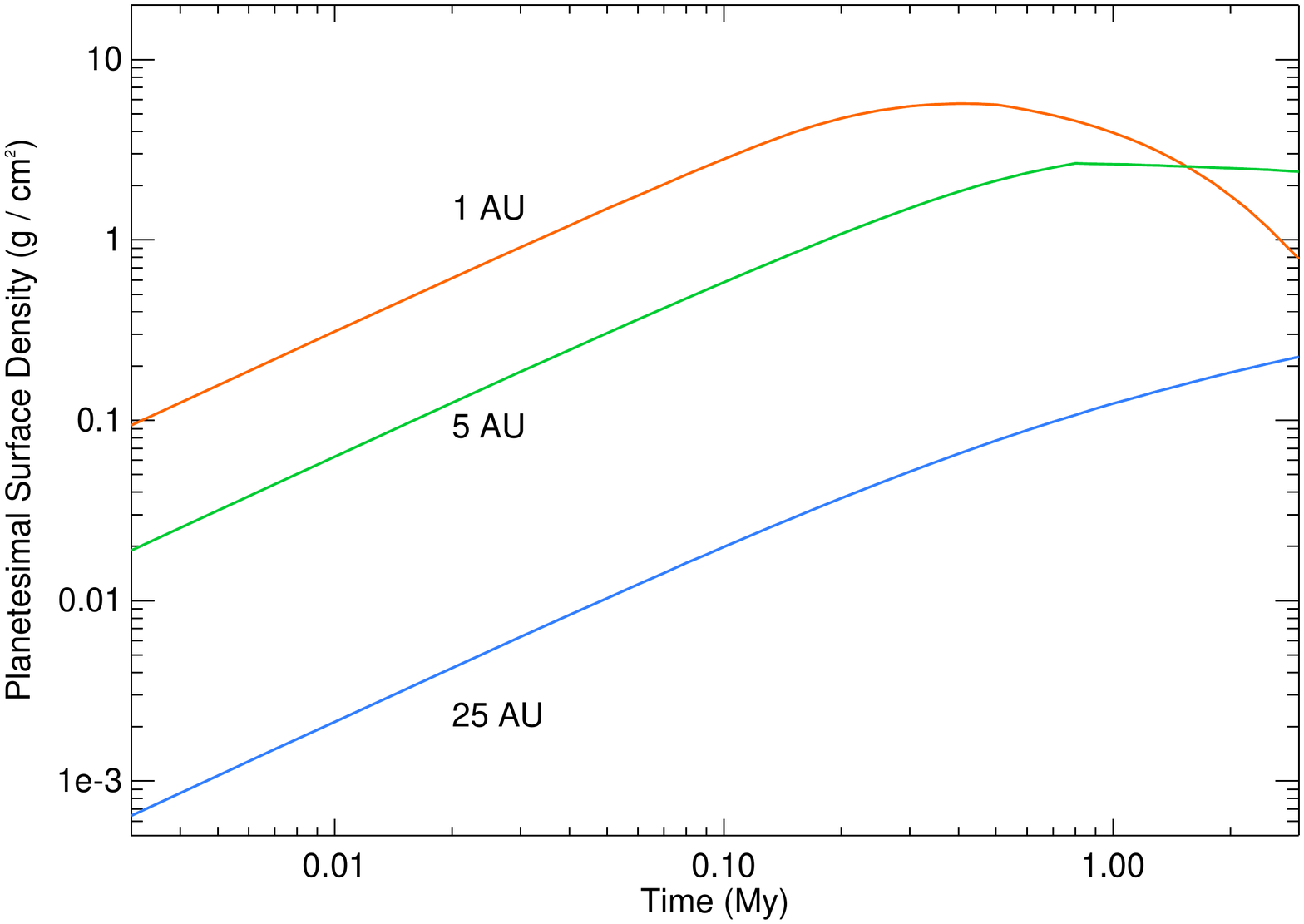}
\caption{Surface density of planetesimals at 3 locations versus time in the simulation shown in Figure~3. The surface density at 1 AU peaks at 0.4 My when planetesimal accretion exceeds planetesimal formation. Soon after this, the inward flux of pebbles ceases and planetesimal formation stops. Planetesimals form throughout the simulation at 25 AU, and few are accreted by embryos.}
\end{figure}

%
% EMBRYO GROWTH
%
\subsection{Embryo Growth}
The red curves in Figure~3 show the embryo mass as a function of distance. Embryo growth rates are fastest just beyond the ice line. Embryos grow more slowly inside the ice line, and more slowly still in the outer parts of the disk. By 3 My, embryos between 2.5 and 12 AU have reached the pebble-isolation mass and stopped growing apart from a gradual increase due to planetesimal accretion. 

The time required to reach the pebble isolation mass increases with distance from the Sun. The solar nebula also cools over time. As a result, the final embryo mass increases more slowly with distance than the instantaneous pebble isolation mass (see Figure 5), and is very roughly 10 Earth masses over a range of distances.

Figure 7 shows the growth histories of embryos at 4 particular locations: 0.4, 1, 5 and 15 AU from the Sun. Growth is initially fastest at 1 AU, with embryos reaching lunar mass by 0.5 My. Growth slows noticeably at this point as pebble accretion ceases. 

At 5 AU, embryos initially grow quite slowly. The growth rate increases rapidly, becoming especially fast between 0.2 and 0.3 My. These objects reach an Earth mass by 0.44 My, and 10 Earth masses by 0.8 My. At this point, pebble accretion ceases and growth essentially stalls. At 15 AU, embryos follow a similar growth trajectory to those at 5 AU, but on a timescale that is an order of magnitude longer.

At 0.4 AU, unlike the other locations, embryo formation is delayed until 0.4 My. Once formed, embryos grow rapidly due to both pebble and planetesimal accretion. Pebble accretion ceases at 0.5 My, but the embryos continue to grow quickly due to planetesimal accretion alone. The reservoir of planetesimals is soon depleted, however, and with no source of new planetesimals the growth rate slows dramatically after 1 My.

\begin{figure}
\plotone{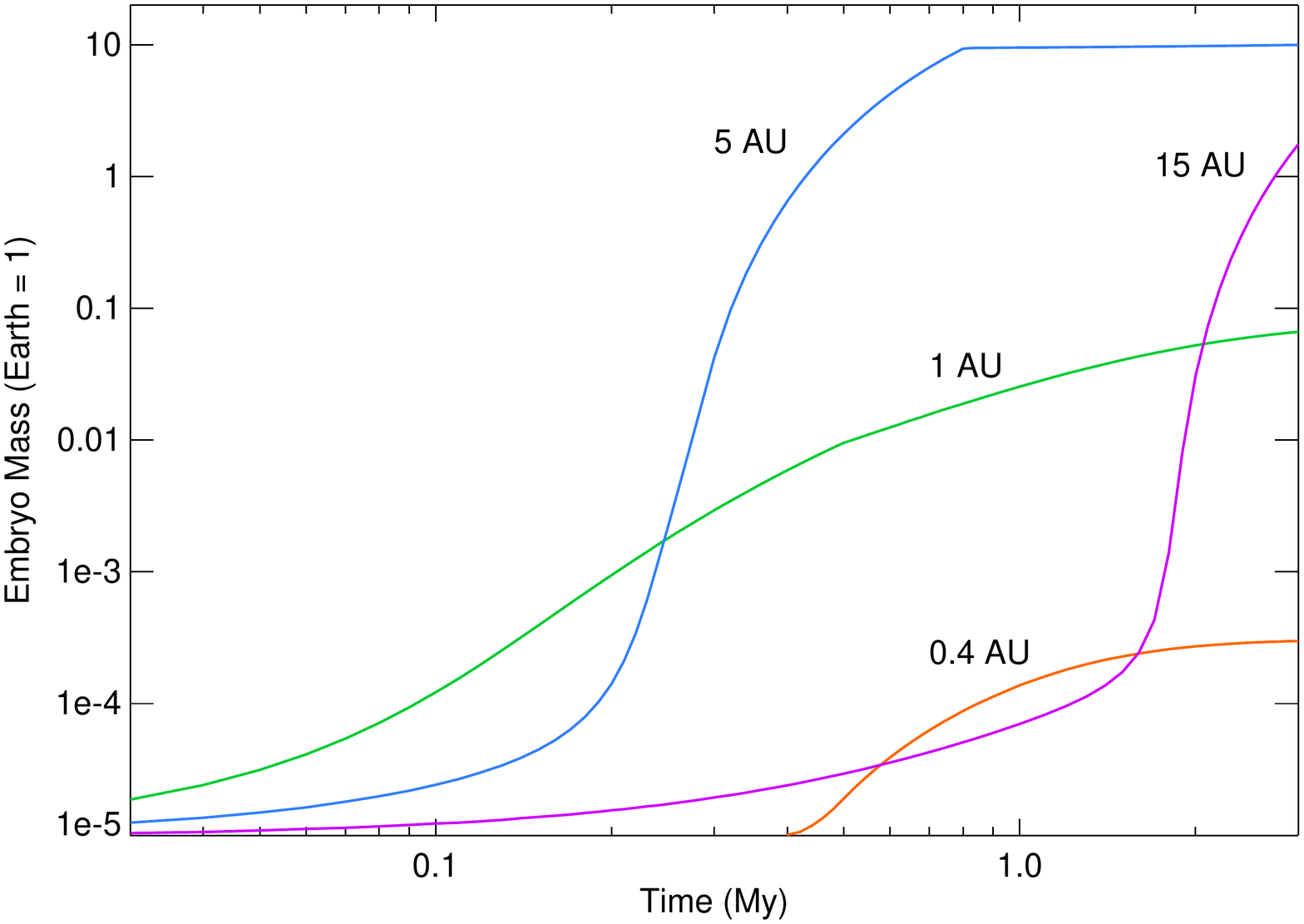}
\caption{Growth of planetary embryos at four locations in the solar nebula in the simulation shown in Figure~3. Embryos are present initially at the three outer locations. At 0.4 AU, an embryo forms after about 0.4 My. The embryo at 5 AU reaches the pebble isolation mass at 0.8 My.}
\end{figure}

A key feature of the Solar System is the small masses of the terrestrial planets compared to the giant planets. This is reproduced in the growth histories of the embryos at 1 and 5 AU in Figure~7, an effect that would be magnified further if gas accretion was included. 

We can understand one of the main reasons for this difference by examining the rate of pebble accretion using the expressions in Section~2. For most of the growth history considered here, embryos undergo pebble accretion in the 3D growth regime. In this case, we can express the growth rate as
\begin{equation}
\frac{d\memb}{dt}=\frac{\memb}{\tgrow}
\end{equation}
where
\begin{equation}
\tgrow=
4.3\alphaturb
\left(\frac{\mstar}{\fpeb}\right)
\left(\frac{c_s^4}{\vkep^3\vfrag}\right)
\left(\frac{\upeb}{\udrift}\right)
\exp\left[2\left(\frac{\st}{\stcrit}\right)^{0.65}\right]
\label{eq-tgrow}
\end{equation}
and
\begin{equation}
\stcrit=\min\left[\frac{4\memb}{\mstar\eta^3},\frac{2}{3}\right]
\label{eq-stcrit}
\end{equation}

\begin{figure}
\plotone{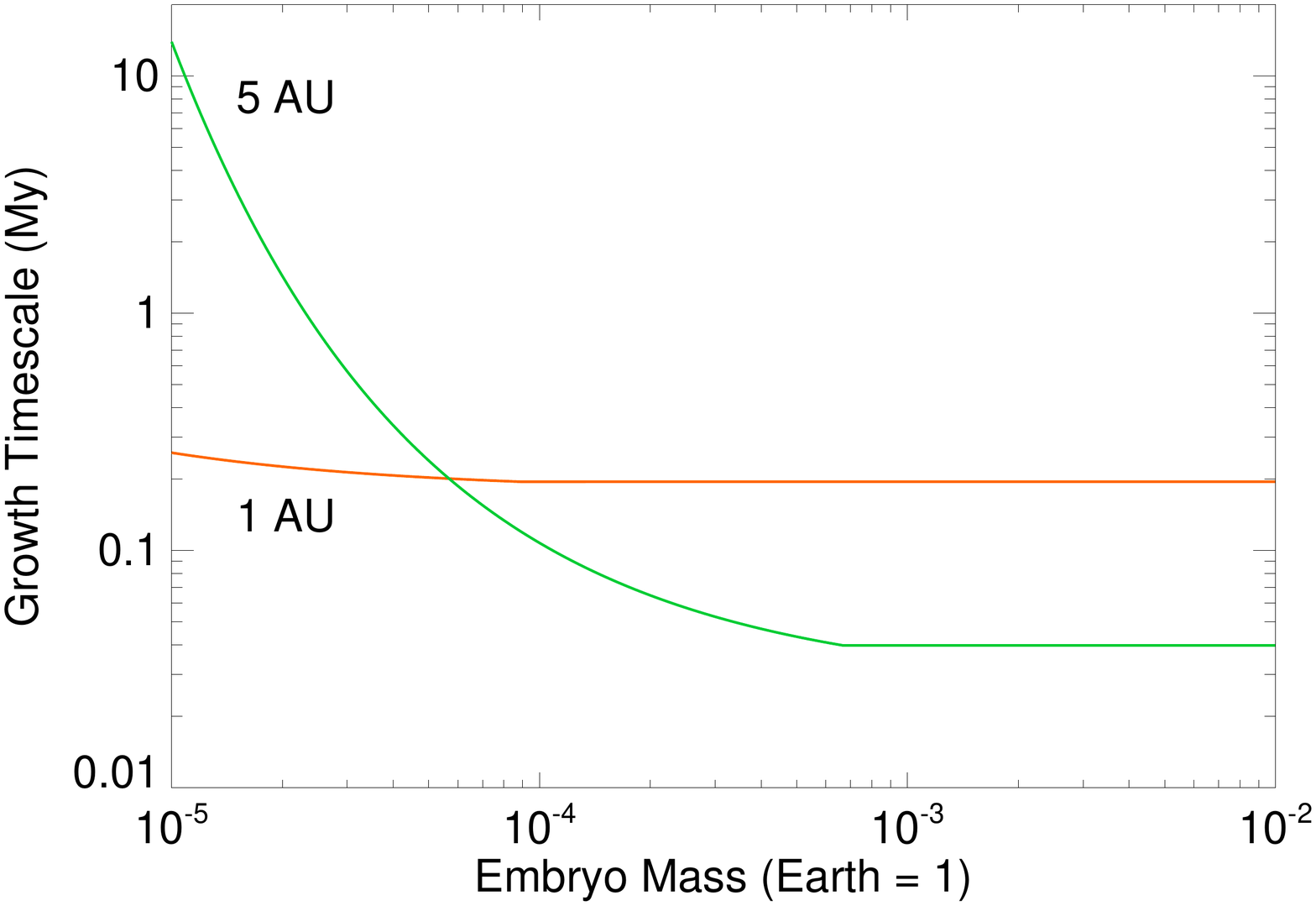}
\caption{Embryo growth timescale as a function of mass at 1 and 5 AU. The pebble flux is $10^{-4}M_\oplus$/y at 5 AU, and a factor of 2 smaller at 1 AU due to ice evaporation.}
\end{figure}

Figure 8 shows $\tgrow$ at 1 AU and 5 AU as a function of the embryo mass. Here we have assumed that $\tmid=230$ and 90 K, respectively, at these locations, which are the values at 0.3 My in the best-fit simulations. The pebble fluxes are taken to be $5\times 10^{-5}M_\oplus$/y and $10^{-4}M_\oplus$/y at 1 AU and 5 AU, respectively, where the difference is caused by ice evaporation at the ice line.

At 1 AU, the growth timescale is roughly $2\times 10^5$ years, almost independent of embryo mass. At 5 AU, $\tgrow$ is much longer than this for the initial embryo mass of $10^{-5}M_\oplus$. However, $\tgrow$ declines rapidly as $\memb$ increases. For $\memb>10^{-3}M_\oplus$, growth is four times faster at 5~AU than at 1~AU.
 
The strong dependence of $\tgrow$ on $\memb$ at 5 AU is caused by the exponential term in Eqn.~\ref{eq-tgrow}, and the fact that the critical Stokes number $\stcrit$ for effective pebble accretion depends on the embryo mass. When the simulation begins, $\st\gg\stcrit$ at 5 AU, and growth is very slow. As the embryos grow, $\stcrit$ increases until the exponential term is no longer significant.

At 1 AU, the embryos are always massive enough that $\st<\stcrit$, and the exponential term in Eqn.~\ref{eq-tgrow} is less important. The notable difference in behavior from that at 5~AU is attributable to the strong radial dependence of the $1/\eta^3$ factor in Eqn.~\ref{eq-stcrit}.

Several factors in Eqn.~\ref{eq-tgrow} favor a shorter  growth timescale at 5 AU than 1 AU when the exponential term can be neglected: (i) the pebble flux is twice as high at 5 AU, (ii) the fragmentation speed is also about 2 times higher, which helps explain why the optimization scheme favors $\vice>\vroc$, (iii) the quantity $\upeb/\udrift$ is about 3 times smaller (this can be seen in Figure~4), and (iv) the sound speed is also smaller. The combination of all these factors outweighs the dependence on $\vkep$ which favors more rapid growth at 1 AU.

We note that an important part of the reason why embryos grow faster at 5 AU than at 1 AU is the contribution of inward gas advection to the radial velocity of pebbles at 1 AU. Pebbles at 1 AU are swept inwards more rapidly by the gas than they would be solely due to drift caused by the pressure gradient. As a result, the pebble surface density at 1 AU is substantially reduced, and growth is corresponding slower.

This effect can be seen in Figure 9, which shows simulations that include (red curves) or ignore (green curves) the contribution of gas advection to the pebble radial velocity. All other model parameters are the same as the case shown in Figure~3. When gas advection is accounted for, embryos inside the ice line are roughly an order of magnitude less massive than they would be otherwise.

\begin{figure}
\plotone{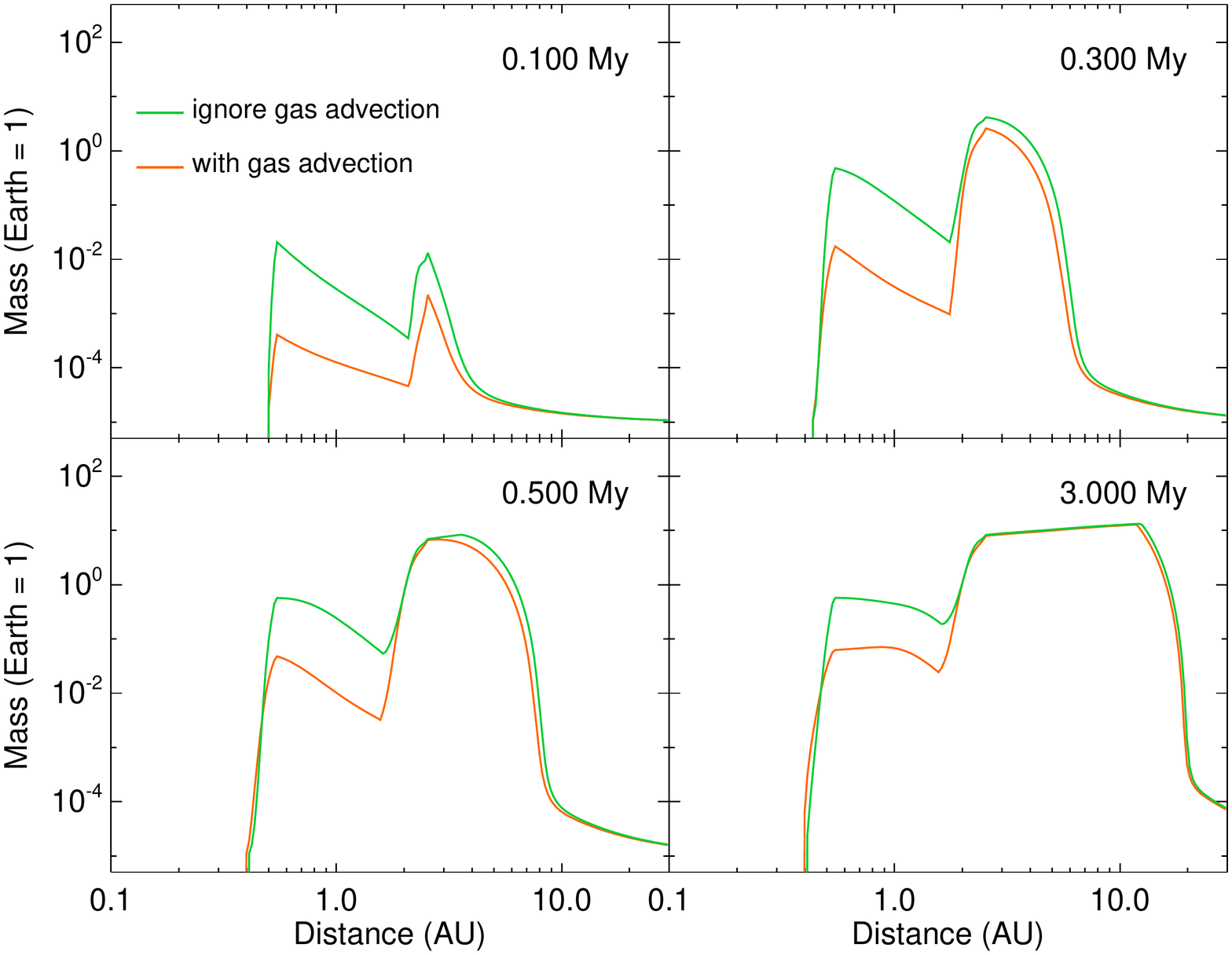}
\caption{Embryo masses as a function of distance. The red curves show a case where the pebble radial velocity includes the effects of gas advection. The green curves show a case in which gas advection is ignored. Model parameters are identical to the case shown in Figure~3.}
\end{figure}

%
% PEBBLE FILTERING
%
\subsection{Pebble Filtering}
An additional effect that limits the growth of embryos in the inner disk is the removal of inflowing pebbles before they reach this region. This ``pebble filtering factor'' is partly the result of pebble accretion by embryos \citep{guillot:2014}. A second, less appreciated aspect is the filtering of pebbles due to the ongoing conversion of pebbles into planetesimals.

We can see the effects of pebble filtering in Figure~10, where the solid curves show the pebble flux as a function of distance at 0.1, 0.3 and 0.6 My (blue, green and red curves respectively). 

At 0.1 My, embryos remain small throughout the disk, and pebble accretion has little effect on the pebble flux. The flux declines slowly with decreasing distance, which is almost entirely due to planetesimal formation. There is also an abrupt decrease by a factor of 2 at the ice line. The flux becomes constant inside 0.5 AU due to the absence of embryos and planetesimal formation.

At 0.3 My, the behavior outside 5 AU is similar to the earlier epoch. The only difference is that the pebble flux entering the planet-forming region has declined by about 40\%. Between 5 and 2 AU, however, the situation changes dramatically. The pebble flux declines by a factor of 3, due largely to efficient pebble accretion by embryos with masses comparable to Earth. Note that this pebble filtering factor acts to starve the inner disk of pebbles, and reduces the pebble accretion rates in the terrestrial-planet region beyond the factors considered in the previous section.

The red curve in Figure 10 shows the pebble flux at 0.6 My. At this stage, embryos at about 3 AU have reached the pebble isolation mass, truncating the flow of pebbles to regions closer to the Sun. At larger distances, the peb flux resembles the situation at 0.3 My, albeit with a further reduction due to the decreased overall pebble flux.

For comparison, the dotted curves in Figure 10 show a simulation in which there is no planetesimal formation. In this case, the pebble flux is essentially independent of distance apart from the discontinuity at the ice line, and the region just outside the ice line where efficient pebble accretion takes place at 0.3 and 0.6 My.

\begin{figure}
\plotone{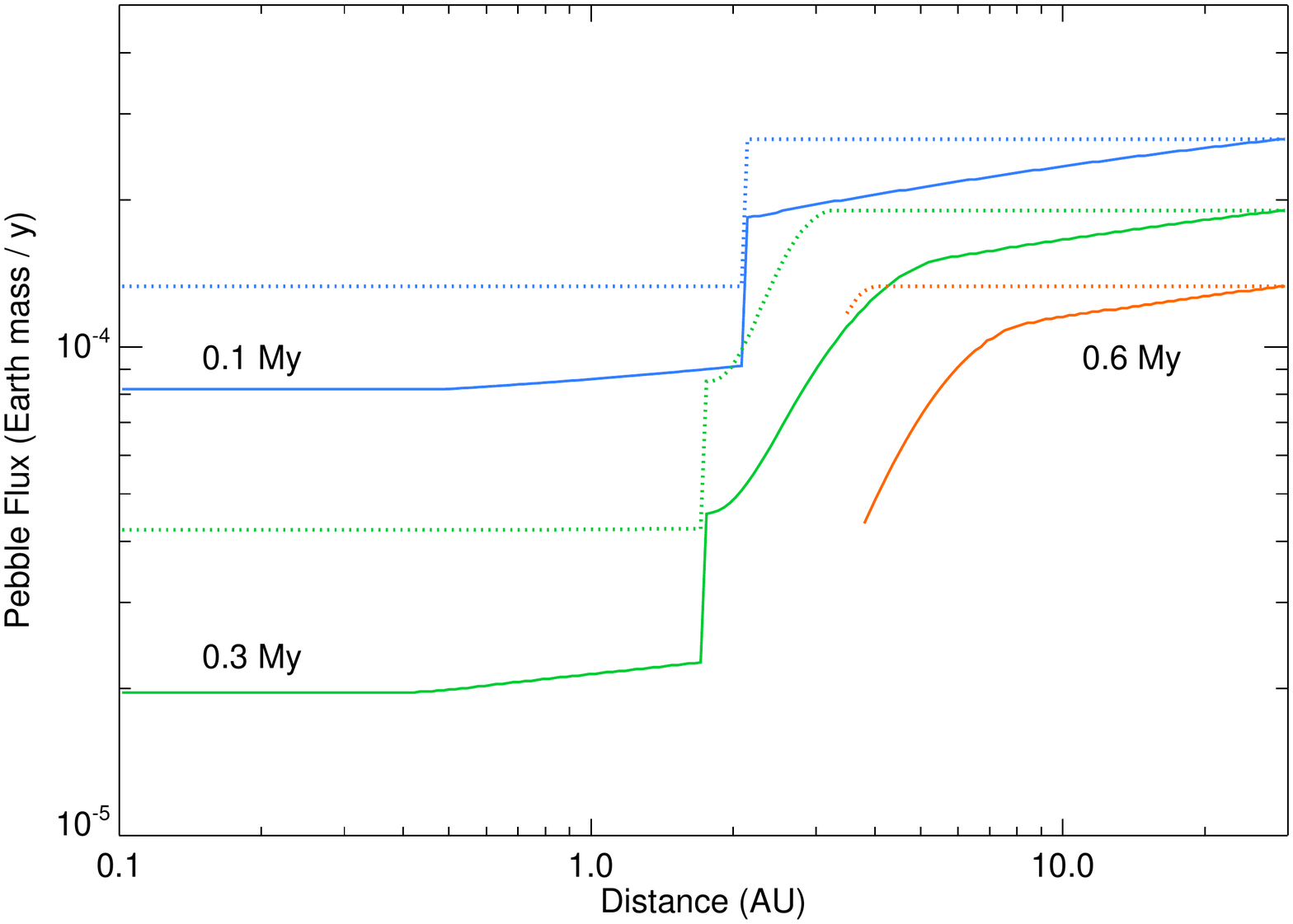}
\caption{Solid curves: pebble mass flux versus distance at 3 times in the simulation shown in Figure~3. Dotted curves: the equivalent fluxes for a simulation without planetesimal formation.}
\end{figure}

%
% PEBBLE VERSUS PLANETESIMAL ACCRETION
%
\subsection{Pebble versus Planetesimal Accretion}
Pebble accretion and planetesimal accretion onto embryos both play a role in planet formation but their relative importance varies with time and location. For example, Figure 11 shows the cumulative mass fraction accreted  in the form of pebbles by embryos. The green and red curves show the pebble contributions after 0.5 and 3 My, respectively.

At both epochs, the disk can be divided into 3 regions based on the pebble fraction of the embryos. The embryos interior to about 1.5 AU have accreted less than half their mass from pebbles, although the pebble contribution is significant, typically at least 10\%. Beyond 1.5 AU, the pebble contribution rises steeply until pebbles dominate the embryo mass budget. Further out, the pebble contribution falls rapidly, at a distance that increases over time. Thus, planetesimals are the dominant component in the outermost disk, although we note that embryos in this region have undergone 
little growth overall (see Figure~3).

The rapid increase in pebble fraction at 1.5 AU roughly coincides with the location of the ice line at 0.5 My. This is also the time when the inward flux of pebbles is first halted by the presence of large embryos further out. Pebble accretion is much more effective beyond the ice line, at this stage, for reasons we described above. Conversely, planetesimal formation tends to become less effective with increasing distance. The transition is smoothed somewhat due to the inward motion of the ice line at earlier epochs. Nonetheless, the transition occurs over a short range of distances.

At 3 My, the transition from mostly planetesimal accretion to pebble-dominated accretion occurs at nearly the same location as at 0.5 My. However, the planetesimal contribution has increased for all embryos out to about 7 AU. This is the result of ongoing planetesimal accretion during the interval, whereas pebble accretion has stopped due to the absence of inward drifting pebbles. Embryos in the region 12--20 AU are still growing at this point, having not yet reached the pebble isolation mass. Pebbles are the dominant component in this region.

We note that the pebble contribution is likely to decrease if residual planetesimals are swept up after the solar nebula has dispersed. This is likely to be particularly important for the terrestrial-planet region. For example, if we assume that all remaining planetesimals in this region are accreted, then the total pebble contribution falls below about 5\% for the region between 1 and 1.5 AU.

\begin{figure}
\plotone{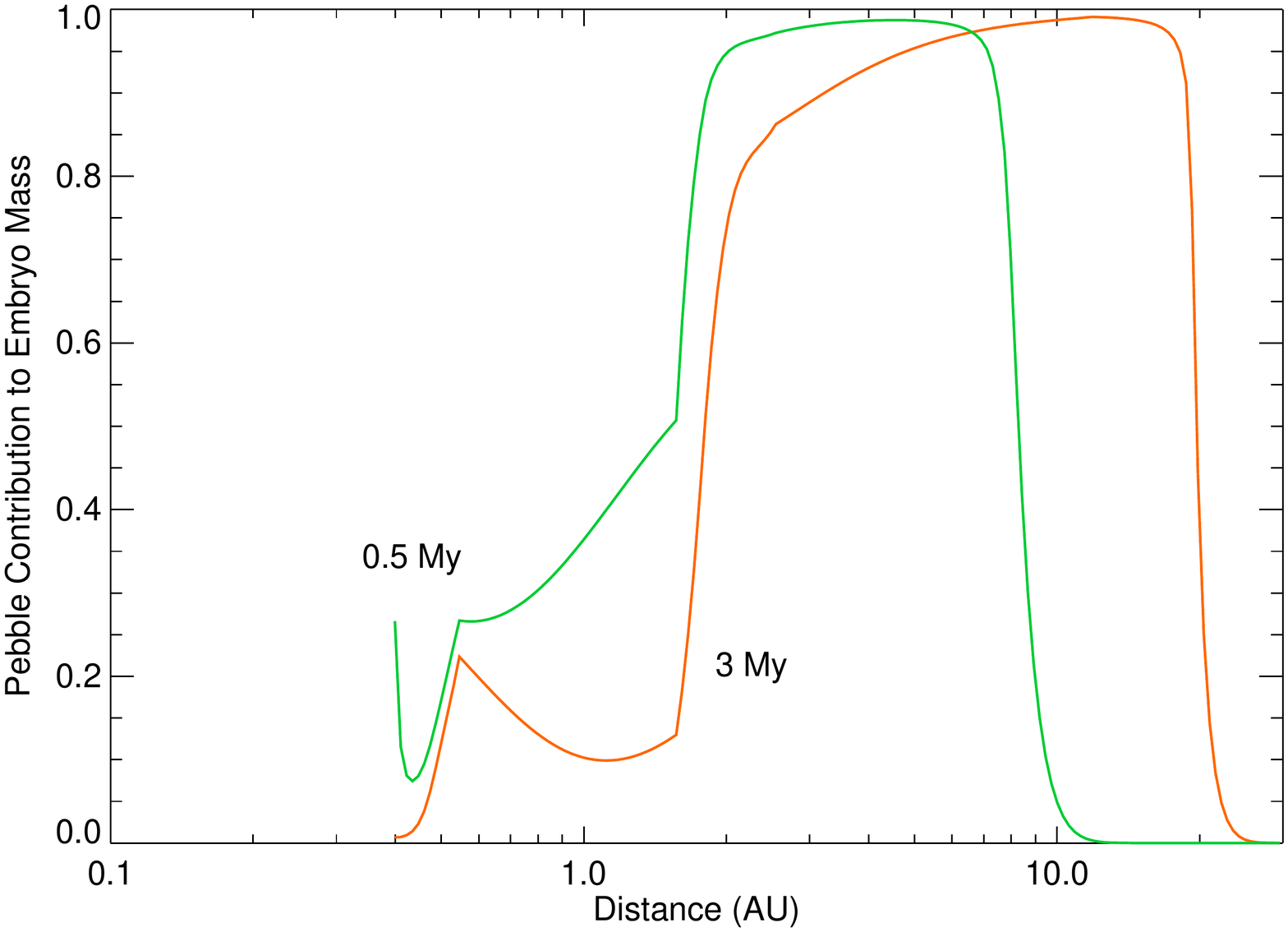}
\caption{Cumulative contribution of pebbles to the total mass of an embryo, as a function of distance, for the simulation shown in Figure~3. The green and red curves shows the situation at 0.5 and 3 My respectively.}
\end{figure}

We can see the role of planetesimal accretion by comparing the red and green curves in Figure~12 which shows  two simulations using identical parameters with and without planetesimal accretion respectively. In the latter case, the embryos can grow only by pebble accretion.

As expected, embryo growth is slower when planetesimal accretion is turned off. The difference varies with distance but is apparent throughout the disk. It makes sense that planetesimal accretion should lead to faster embryo growth in the inner disk where planetesimal accretion is quite efficient. However, embryos are also much larger, or grow much faster in the outer disk.

The reason for this behavior becomes clearer by examining the green curve in the final panel of Figure~12. Typically embryos beyond the ice line have either reached the pebble isolation mass (roughly 2--6 AU), or they remain close to their initial mass ($>6$ AU). 

As we saw earlier, pebble accretion is largely ineffective when $\st\gg\stcrit$. This is the case for all initial embryos outside the ice line. Initially therefore, growth by pebble accretion is very slow, and it remains so until the embryos have grown large enough that $\stcrit\sim\st$. (Recall from Eqn.~\ref{eq-stcrit} that $\stcrit\propto\memb$.) Once this happens, embryos grow very rapidly by pebble accretion until they reach the pebble isolation mass.

When planetesimal accretion is included, significant early growth occurs despite the ineffectiveness of pebble accretion. This demonstrates an important role of planetesimal accretion outside the ice line, namely to give an early boost to embryo masses, allowing them to reach the point where rapid pebble accretion can take over. This allows embryos to reach the pebble isolation mass over a much larger region of the disk by the end of the simulation. Thus, early growth due to the accretion of planetesimals leads to much more effective formation of giant-planet cores by triggering rapid pebble accretion.

\begin{figure}
\plotone{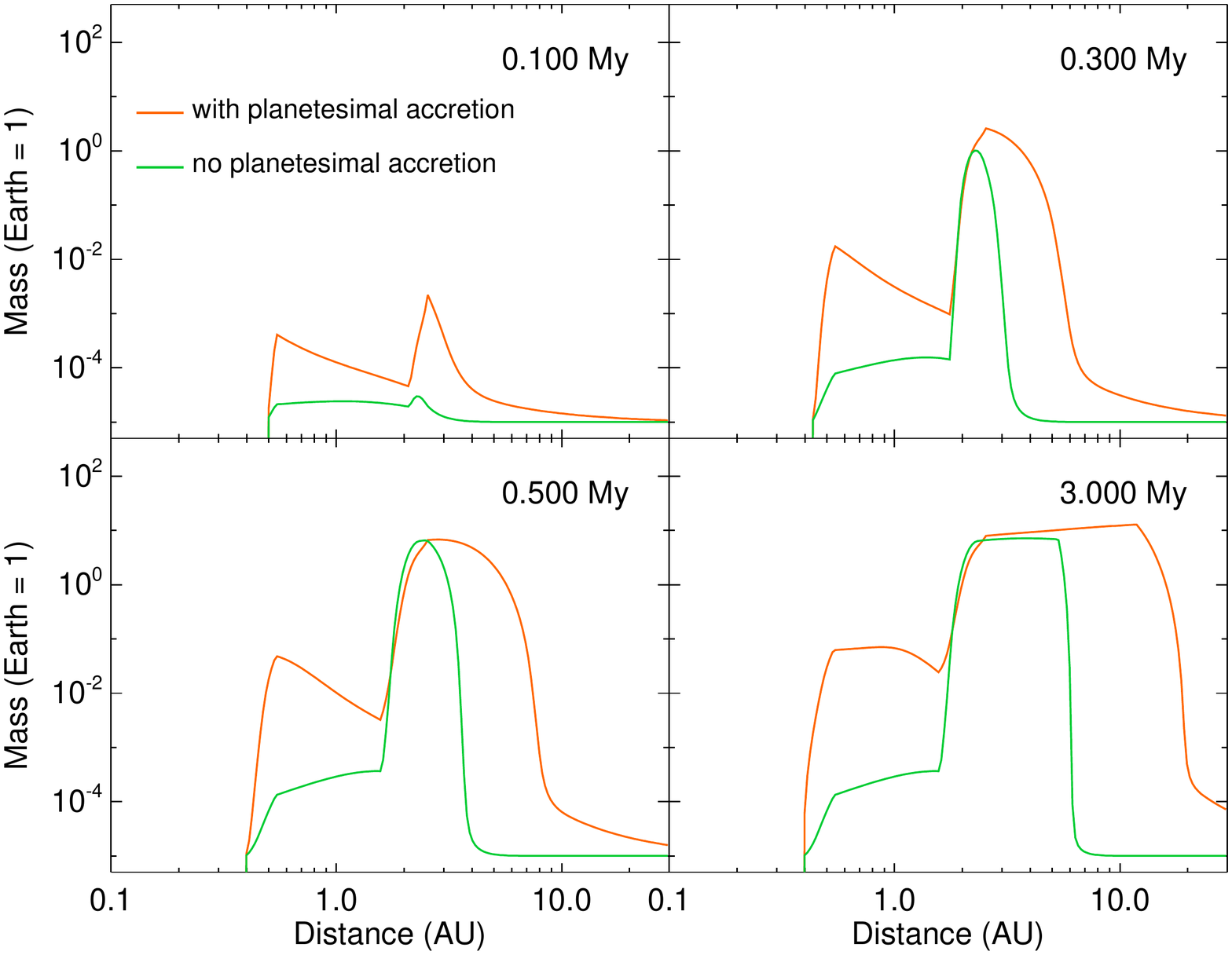}
\caption{Embryo masses as a function of distance in simulations with (red curves), and without (green curves) planetesimal accretion. Model parameters are identical to the case shown in Figure~3.}
\end{figure}

%
% VARYING THE MODEL PARAMETERS
%
\section{Varying The Model Parameters}
As we saw in the previous sections, the evolution can be quite complex even in the relatively simple planet formation model used here. The evolution depends on the 6 poorly known parameters identified in Section 3. In one or two cases, the effect of a parameter is reasonably clear. For example, $\tauwind$ is the only parameter that affects the disk temperature. However, most aspects of the evolution depend on several parameters in a more complicated manner. In the following sections, we aim to understand this behavior by varying each parameter individually.

\begin{figure}
\plotone{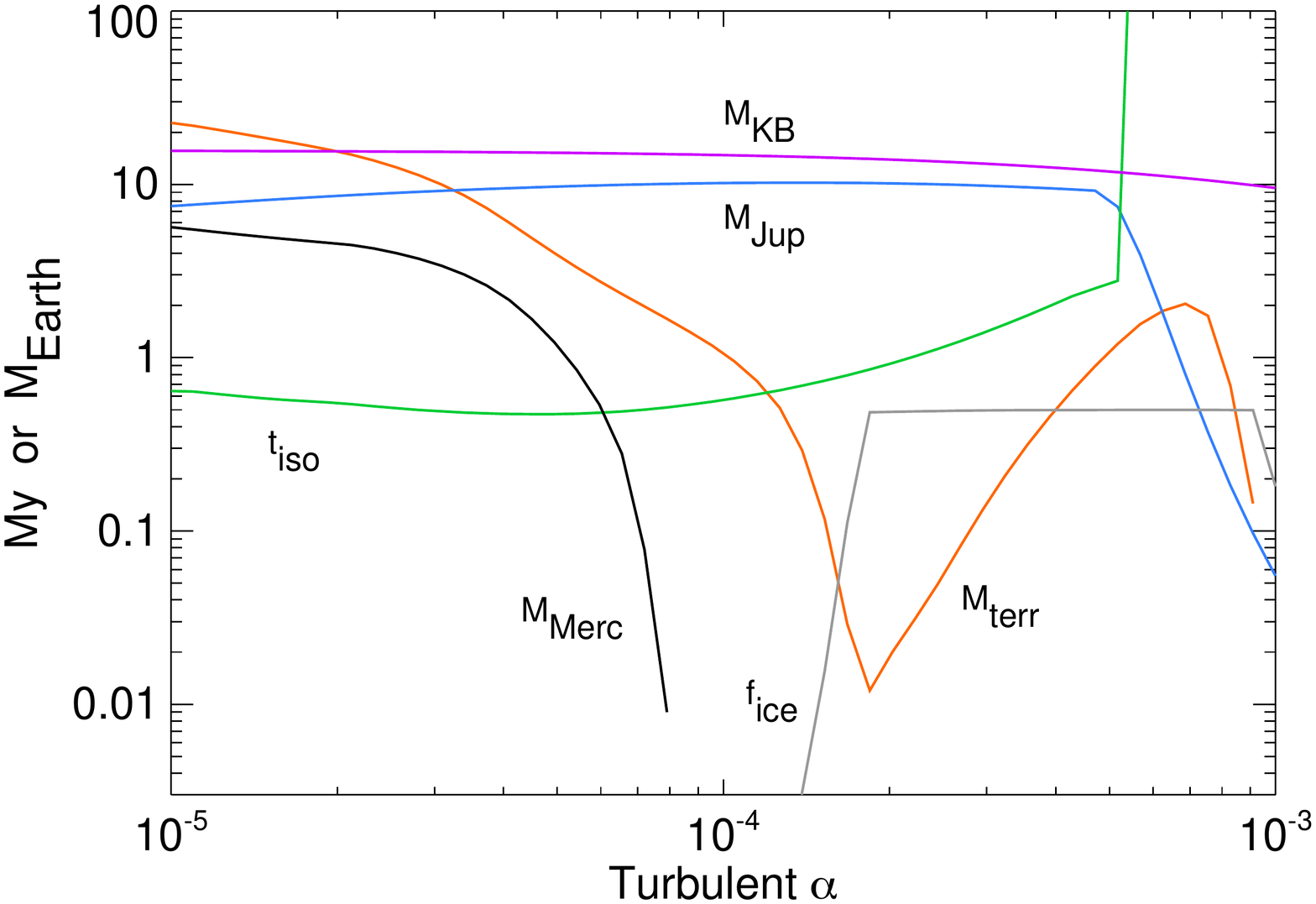}
\caption{The effect of varying the turbulence strength $\alphaturb$ on the Solar System constraints. Other model parameters have the same values as the case shown in Figure~3.}
\end{figure}

%
% TURBULENCE STRENGTH
%
\subsection{Turbulence Strength $\alphaturb$}
Figure 13 shows the effect of varying the turbulence strength $\alphaturb$ on the values of the Solar System constraints. All other parameters have the same values as the case shown in Figure~3. Changing the turbulence strength affects the evolution by altering pebble accretion in two ways. Firstly, larger $\alphaturb$ leads to more energetic pebble collisions, and thus smaller pebbles and smaller $\st$. Secondly, stronger turbulence increases the pebble scale height $\hpeb$, reducing the space density of pebbles.

Interestingly, the embryo mass at 5 AU, $\mjup$, is largely unaffected by changes in $\alphaturb$ for values $<4\times 10^{-4}$. These embryos all reach the pebble isolation mass $\miso$ within 3 My and essentially stop growing. $\miso$ itself depends very weakly on $\alphaturb$, so the final embryo mass at 5 AU is roughly constant. 

However, raising $\alphaturb$ typically results in slower growth at 5 AU, leading to larger values of the pebble isolation time $\tiso$, at least for $\alphaturb>5\times 10^{-5}$. When $\alphaturb>5\times 10^{-4}$, embryos take more than 3 My to reach the pebble isolation mass. For these embryos, the growth timescale $\tgrow\propto\alphaturb^2$ in Eq.~\ref{eq-tgrow} because gas advection dominates the radial motion of pebbles, so that $\upeb/\udrift\propto1/\alphaturb$. As a result, $\mjup$ declines steeply with further increases in $\alphaturb$.

We note that for $\alphaturb<5\times 10^{-5}$, the growth rate at 5 AU is nearly constant. In this case, the weak turbulence and large values of $\st$ lead to very small values of $\hpeb$. Much of the growth of embryos at 5 AU takes place in the 2D growth regime as a result, which is slower than the equivalent 3D growth rate would be for $\rcap<\hpeb$. This effect counteracts the nominally faster growth caused by larger $\st$.

The timing of pebble isolation also determines the ice fraction $\fice$ of the terrestrial planets. When $\alphaturb$ is small, pebble isolation occurs before the ice line enters the terrestrial planet region, leading to $\fice=0$. Stronger turbulence delays the onset of pebble isolation so that icy pebbles can enter the terrestrial planet region, leading to larger $\fice$. When $\alphaturb>2\times 10^{-4}$, solid material in the terrestrial planet region becomes almost 50\% ice by mass.

The total mass of material $\mterr$ inside 1.5 AU depends on $\alphaturb$ in a complicated manner. For the smallest values of $\alphaturb$, we have $\st>\stmin$ throughout the disk so that embryos can form everywhere. As turbulence increases, the region where embryos can form shrinks, reducing the extent of the terrestrial planet region. This is reflected in the decline in the mass $\mmerc$ interior to 0.5 AU

In addition, larger $\alphaturb$ reduces the efficiency of pebble accretion in the inner disk, reducing embryo growth. Growth in the terrestrial region slows down with the onset of pebble isolation because only planetesimal accretion can continue at this point. As we saw above, the timing of pebble isolation, $\tiso$, is nearly constant for $\alphaturb<5\times 10^{-5}$. Thus, the net effect of raising $\alphaturb$ is to reduce the mass of the terrestrial planets with little change for embryos at 5 AU.

When $\alphaturb>10^{-4}$, we see that $\mterr$ decreases rapidly. At this point, the inner edge of the region where embryos can form is approaching the outer edge of the terrestrial planet region at 1.5 AU. The situation changes abruptly when $\alphaturb$ exceeds $2\times 10^{-4}$. Now, the increase in $\tiso$ means that icy pebbles can enter the region inside 1.5 AU. These pebbles have larger $\st$, allowing embryos to form here, albeit quite late in the evolution. Further increases in $\alphaturb$ delay pebble isolation even more, allowing $\mterr$ to increase despite the reduced efficiency of pebble accretion.

Finally, we note that the mass $\mkb$ of planetesimals forming in the proto-Kuiper belt is essentially independent of $\alphaturb$. The planetesimal formation rate depends only on the pebble flux and  $\hgas$ provided that $\upeb\simeq\udrift$. This is true in the outermost disk for all but the largest values of $\alphaturb$ considered here.

\begin{figure}
\plotone{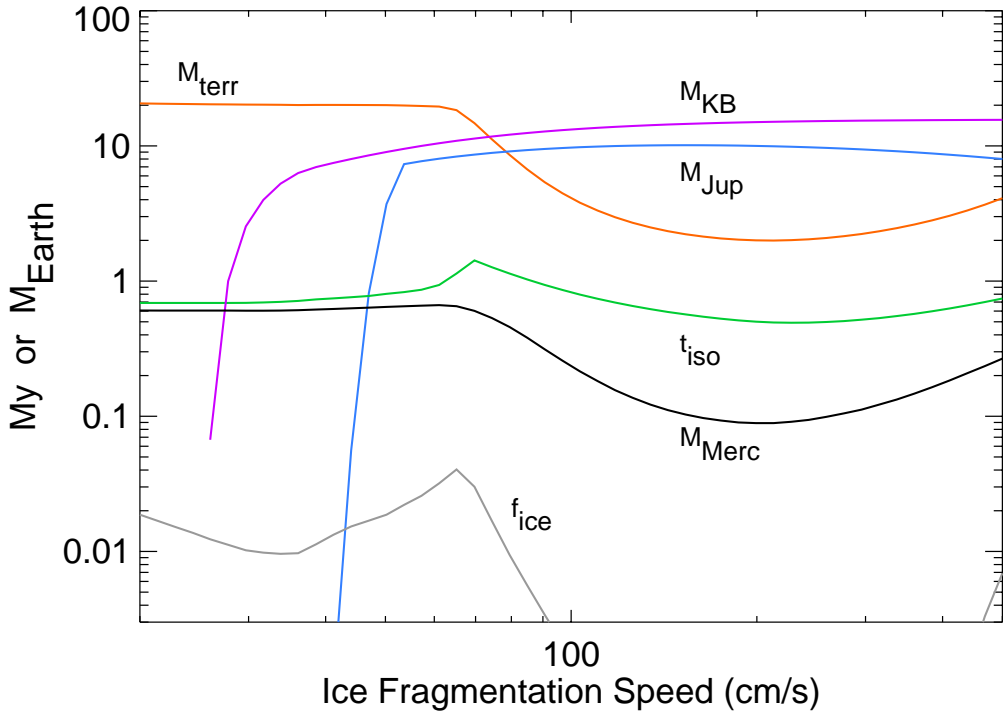}
\caption{The effect of varying the ice fragmentation velocity $\vice$ on the Solar System constraints. Other model parameters have the same values as the best-fit case shown in Figure~3.}
\end{figure}

%
% ICE FRAGMENTATION VELOCITY
%
\subsection{Ice Fragmentation Velocity $\vice$}
The ice fragmentation speed $\vice$ is one of the two main factors that determine the pebble Stokes number beyond the ice line, the other being the turbulence strength. Thus $\vice$ affects whether embryos and planetesimals can form (which requires $\st>\stmin$), and also affects the pebble accretion rate onto embryos. Figure~14 shows the effect of varying $\vice$, with the other parameters held fixed.

Small values of $\vice$ lead to very small $\st$, so that embryos cannot form in many parts of the disk. In extreme cases, when $\vice<25$ cm/s, we find that embryos are unable to form in the Kuiper-belt region (here taken to be $>20$ AU). For $\vice<40$ cm/s embryos cannot form at 5 AU. This is the reason why $\mkb$ and $\mjup$, respectively, fall to zero when $\vice$ is small.

As the ice fragmentation speed increases beyond about 70 cm/s, larger pebbles are able to exist, and this leads to more effective pebble accretion. As a result, the pebble isolation time decreases. Once $\vice$ exceeds about 230 cm/s, however, $\tiso$ starts to increase again. At this point, the pebble scale height has become small enough that much of the growth of embryos beyond the ice line takes place in the 2D regime, which is slower than the equivalent 3D growth rate for $\rcap<\hpeb$. From this point on, the reduced pebble surface density due to higher drift rates becomes the more important factor, and growth rates decline with increasing $\vice$. There is a broad ``sweet spot'' for low values of $\tiso$ with $\vice\sim 200$ cm/s, which explains why the best-fit simulations tend to lie in this part of phase space.

Inside the ice line, the formation of embryos and planetesimals is controlled by $\vroc$ rather than $\vice$, so bodies can form in this region even if they are prevented from forming beyond the ice line. However, growth in the terrestrial-planet region is affected by $\vice$ indirectly because of its influence on the timing of pebble isolation $\tiso$.

For $\vice>65$ cm/s, the solid mass $\mterr$ in the terrestrial region is clearly controlled by $\tiso$. Larger $\tiso$ allows more time for embryos in the terrestrial zone to grow by pebble accretion, leading to a larger total mass in the this region. The mass $\mmerc$ interior to 0.5 AU closely tracks the behavior of $\mterr$ since it is controlled by the same factors.

For $\vice<65$ cm/s, $\mterr$ is essentially constant. In this case, some embryos in the terrestrial planet region are able to reach the pebble isolation mass (roughly $2.5M_\oplus$ at 1 AU) before the end of the simulation. This shuts off the supply of pebbles, halting pebble accretion and preventing any more planetesimals from forming. The remaining planetesimals can still be accreted, but this doesn't change the total mass in the region.

The fact that embryos inside the ice line are able to reach pebble isolation explains why $\tiso$ becomes roughly constant for small $\vice$ despite large variations in the behavior beyond the ice line. In this case, the first (and often only) embryos to reach pebble isolation lie inside the ice line.

\begin{figure}
\plotone{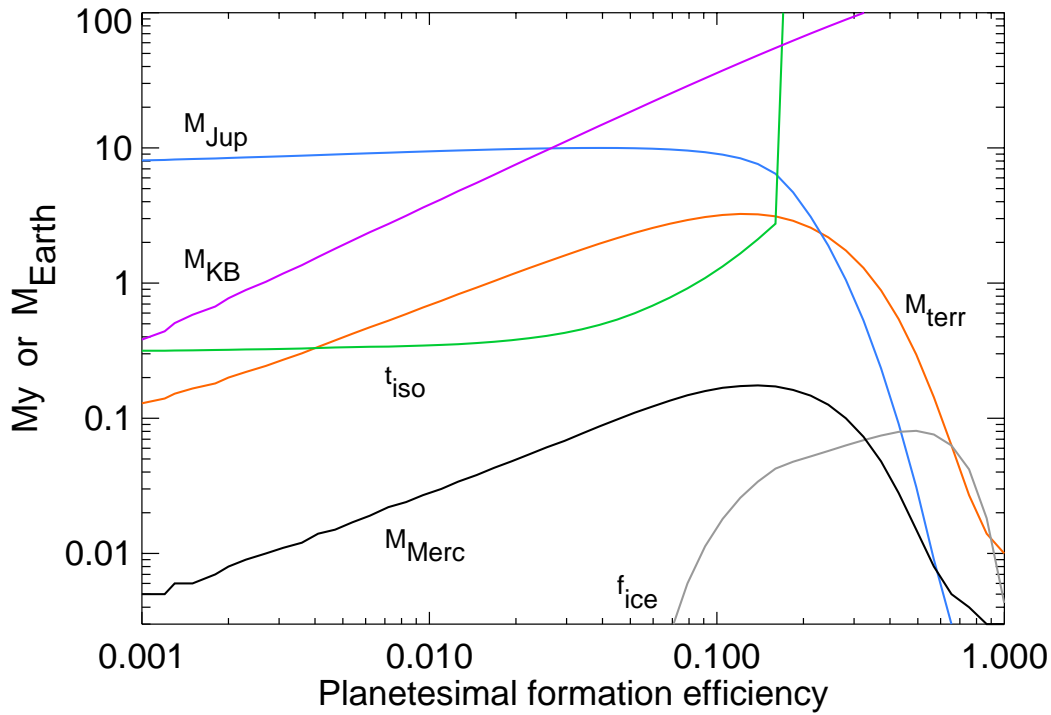}
\caption{The effect of varying the planetesimal formation efficiency $\epsilon$ on the Solar System constraints. Other model parameters have the same values as the best-fit case shown in Figure~3.}
\end{figure}

%
% PLANETESIMAL FORMATION EFFICIENCY
%
\subsection{Planetesimal Formation Efficiency $\epsilon$}
The planetesimal formation efficiency parameter $\epsilon$ controls the rate at which pebbles are converted into planetesimals provided that $\st\ge\stmin$. It also determines the fraction of the pebble flux that {\em doesn't\/} form planetesimals and is thus available for pebble accretion by embryos closer to the Sun. Figure~15 shows the effect of varying $\epsilon$ while the other parameters remain fixed.

When $\epsilon$ is small, it has little effect on embryo growth outside the ice line. Only a small fraction of pebbles are converted into planetesimals, so planetesimal accretion is negligible in this region, and pebble accretion rates are nearly unaffected. Both the embryo mass $\mjup$ at 5 AU and the time $\tiso$ of the onset of pebble isolation are roughly constant as a result.

Larger values of $\epsilon$ lead to longer times until pebble isolation begins. Although more planetesimals are available for planetesimal accretion this is more than offset by the reduction in the pebble flux so that $\tiso$ increases. When $\tiso$ approaches the simulation length of 3 My, the embryos at 5 AU also fail to reach the pebble isolation mass, and $\mjup$ starts to decline rapidly. This behavior is similar to the steep drop off in $\mjup$ for large $\alphaturb$ seen in Figure~13, and for the same reason.

Once $\tiso$ exceeds about 0.6 My, icy pebbles are able to enter the terrestrial-planet region before embryos at larger distances reach pebble isolation. This leads to a significant ice fraction of the solid material in the terrestrial planet region.

For $\epsilon<0.1$, the mass of solid material in the terrestrial region, and the region occupied by Mercury, both increase with increasing $\epsilon$. This occurs for two reasons. Firstly, increased planetesimal formation directly adds to the reservoir of solid material in these regions, even if the planetesimals take time to be accreted by embryos. Secondly, increasing $\epsilon$ delays pebble isolation at larger distances, as we noted above. This allows longer for planetesimal formation and pebble accretion to operate in the inner disk.

However, when $\epsilon$ becomes larger than about 0.15, both these factors become outweighed by the reduced flux of pebbles reaching the inner disk due to planetesimal formation further out. At this point, both $\mmerc$ and $\mterr$ begin to decline rapidly with increasing $\epsilon$.

Finally, as expected, the mass $\mkb$ of planetesimals that form in the proto-Kuiper belt depends linearly on $\epsilon$.

\begin{figure}
\plotone{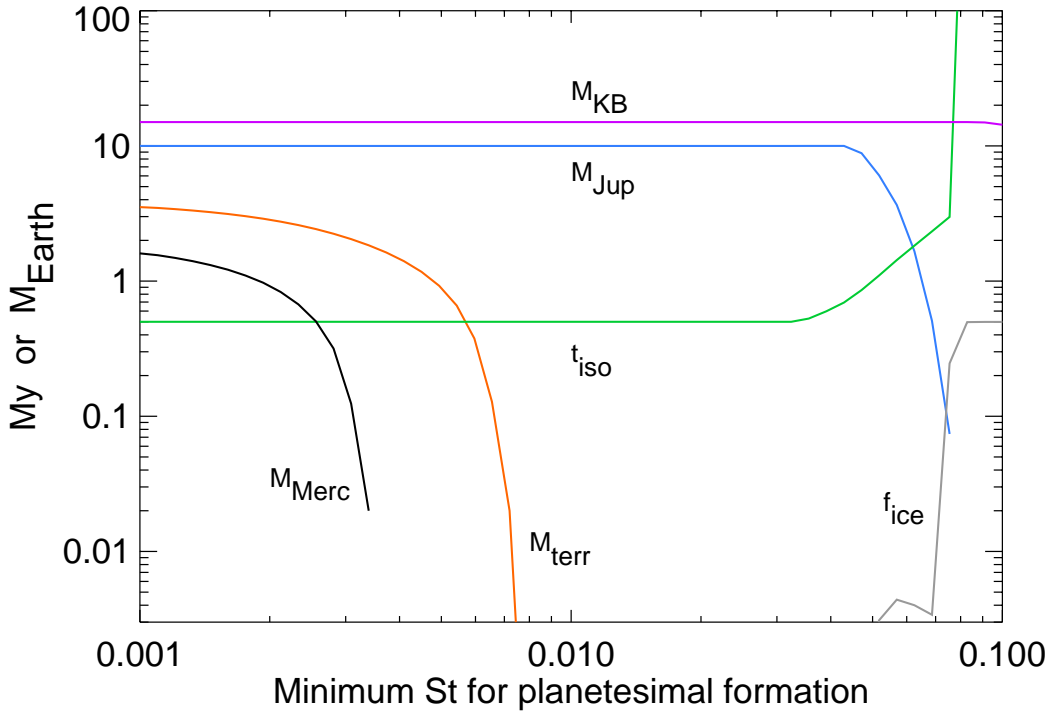}
\caption{The effect of varying the minimum Stokes number $\stmin$ for planetesimal formation on the Solar System constraints. Other model parameters have the same values as the best-fit case shown in Figure~3.}
\end{figure}

%
% MINIMUM STOKES NUMBER FOR PLANETESIMAL FORMATION
%
\subsection{Minimum Stokes Number $\stmin$ for Planetesimal Formation}
In our model, planetesimals and planetary embryos are able to form only when the pebble stokes number $\st$ exceeds $\stmin$. Figure~16 shows the effect of varying $\stmin$ while the other parameters retain their values from the simulation shown in Figure~3.

Beyond the ice line, the main effect of $\stmin$ is to determine where and when embryos can form. Both $\mjup$ and $\tiso$ are nearly constant for $\stmin<0.03$. In this case, embryos are able to form at the start of the simulation and growth proceeds normally. 

For larger values of $\stmin$, embryos cannot form initially at 2.5--5 AU because the disk is too hot. The resulting turbulent velocities are high enough, and pebble collisions are energetic enough, that $\st<\stmin$. Later, when the disk cools, $\st$ increases, and embryos are able to form here. However, the delay means that embryos take longer to reach the pebble isolation mass, or do not reach this mass at all before the simulation ends. This leads to larger $\tiso$ and smaller $\mjup$. When $\stmin$ exceeds about 0.07, embryos are unable to form at 5 AU at any time.

Inside the ice line, the pattern is similar except that embryo and planetesimal formation is prevented at much smaller values of $\stmin$. For example, we require $\stmin<0.0035$ in order for planetesimals and embryos to form in the region occupied by Mercury, otherwise planet formation is prevented here entirely.

In principle, planetesimal formation in the proto-Kuiper belt will also be affected by $\stmin$. However, in all cases considered here, we find that $\st>\stmin$ at the start of the simulation, or soon afterwards. As a result, the mass of planetesimals in this region is essentially independent of $\stmin$.

\begin{figure}
\plotone{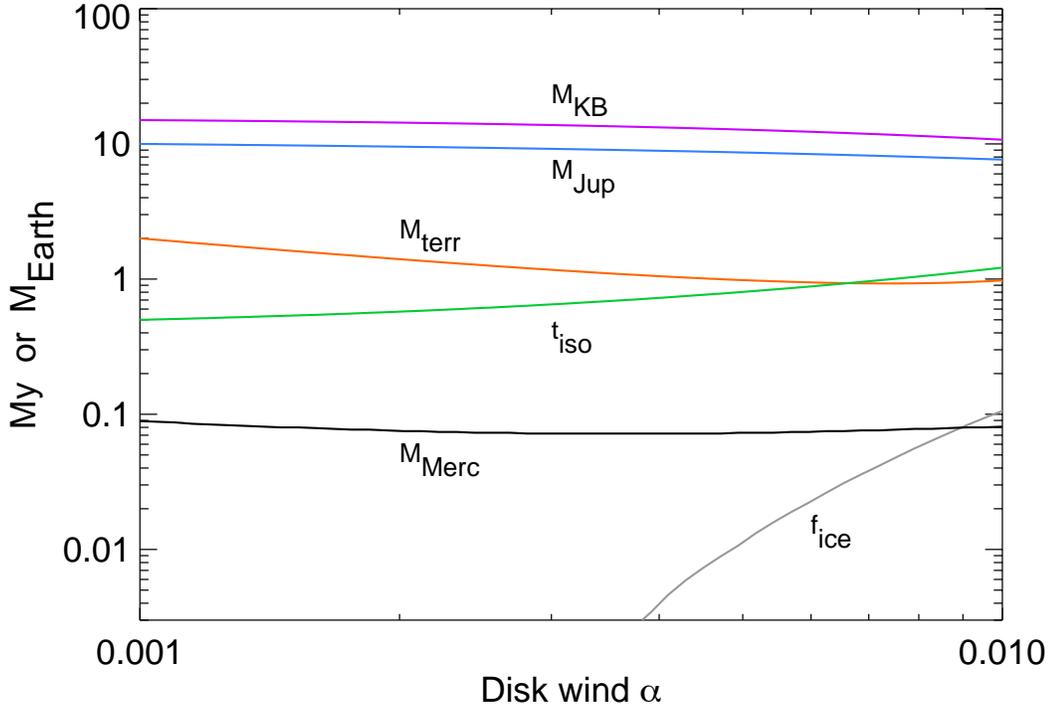}
\caption{The effect of varying the disk wind strength parameter $\alphawind$ on the Solar System constraints. Other model parameters have the same values as the best-fit case shown in Figure~3.}
\end{figure}

%
% DISK WIND STRENGTH
%
\subsection{Disk Wind Strength $\alphawind$}
Figure~17 shows the effect of varying the disk wind parameter $\alphawind$ while the other parameters have the same values as the simulation shown in Figure~3.

The changes caused by varying $\alphawind$ are more subtle than for some of the model parameters considered above. The main role of $\alphawind$ is to determine the inward radial velocity $\ugas$ of the gas, which is a component of the inward velocity $\upeb$ of the pebbles. This in turn affects the surface density of pebbles $\sigmapeb$. Larger values of $\alphawind$ lead to smaller values of $\sigmapeb$, and thus lower rates of planetesimal formation and pebble accretion.

As we saw in Figure 4, $\upeb$ in the outer disk is dominated by radial drift rather than gas advection. For this reason, $\alphawind$ has only a modest effect on $\mjup$ and $\mkb$. The effect on the timing $\tiso$ of pebble isolation is significant however, with $\tiso$ increasing by about a factor of 2 for the range of $\alphawind$ values considered here. As we noted earlier, when $\tiso$ exceeds about 0.5 My, icy pebbles are able to enter the terrestrial-planet region before pebble isolation begins, and this is reflected in the non-zero values of $\fice$ at larger values of $\alphawind$.

One might expect that the delayed onset of pebble isolation would also lead to an increase in the solid mass $\mterr$ in the terrestrial-planet region. However, this effect is more than offset by the decreased efficiency of pebble accretion and planetesimal formation due to the reduction in $\sigmapeb$. As a result, $\mterr$ declines with increasing disk wind strength.

\begin{figure}
\plotone{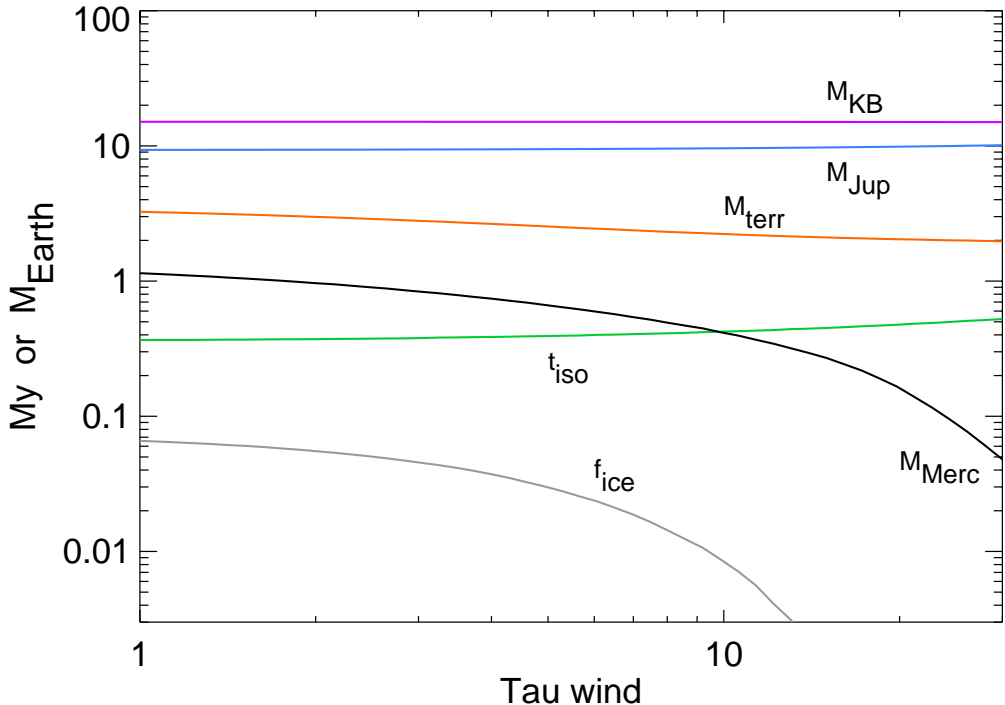}
\caption{The effect of varying the optical depth $\tauwind$ of the heated layer of the disk on the Solar System constraints. Other model parameters have the same values as the best-fit case shown in Figure~3.}
\end{figure}

%
% OPTICAL DEPTH OF THE HEATED LAYER
%
\subsection{Optical Depth of the Heated Layer $\tauwind$}
The disk wind is assumed to heat the disk in a layer with an optical depth of $\tauwind$ from the disk surface. This parameter helps determine the midplane temperature, particularly in the inner part of the disk. It is less in important in the outer disk where the temperature is determined mainly by stellar irradiation. Figure~18 shows the effect of varying $\tauwind$ while other parameter values are held fixed.

An increase in $\tauwind$ leads to a hotter inner disk, reducing the amount of time that icy pebbles can exist here. As a result, the ice fraction of the inner planets declines for larger values of $\tauwind$.

When $\tauwind$ is small, a significant amount of terrestrial-planet region receives icy pebbles before the supply is terminated by pebble isolation. Icy pebbles are stronger than rocky ones for the best-fit value of $\vice$. The pebbles are larger as a result, and pebble accretion proceeds more rapidly. More solid material is available due to the presence of ice. Together, these factors lead to faster growth in the inner disk when $\tauwind$ is small, and $\mterr$ is also larger as a result. 

Because small values of $\tauwind$ lead to larger values of $\st$, we find that planetesimals and embryos are able to form closer to the sun in this case. As a result, the mass $\mmerc$ of solid material in the region occupied by Mercury increases substantially, exceeding an Earth mass when $\tauwind=1$.

%
% DISCUSSION
%
\section{Discussion}

%
% PLANETARY MASSES
%
\subsection{Planetary Masses}
A striking feature of the Solar System is the large difference in mass between the inner and outer planets. Even when the gaseous envelopes of the outer planets are removed, they are      still an order of magnitude more massive than the terrestrial planets.

\citet{morbidelli:2015} used simple growth models to show that planetesimal accretion alone cannot explain the mass difference between the Mars-mass bodies that probably formed in the inner solar nebula and the 10-Earth-mass cores of the giant planets. Instead, this difference could have arisen if these objects formed by pebble accretion provided that the pebble Stokes number dropped by a factor of 10 inside the ice line. They argued that the evaporation of water ice from pebbles drifting inwards across the ice line would have released much smaller rocky grains leading to a sharp drop in pebble size.

The results in Sections 4 and 5 support this conclusion to some extent. However, we find that additional factors are likely to play a role in determining planetary sizes.

The rocky grains released by ice evaporation will presumably collide and stick together, quickly forming larger and larger aggregates until collisions become too energetic. Thus, the pebble size inside the ice line is controlled by the velocity at which collisions change from growth to fragmentation, just as it is outside the ice line. Unfortunately, the fragmentation velocity of water ice under nebular conditions is not known. Some experiments suggest that icy particles can stick at much higher speeds than rocky ones, but other work suggests this is not the case \citep{gundlach:2015, musiolik:2019}. Planet formation studies can help constrain $\vice$. For example, the best-fit simulations in Section~4 predict an ice fragmentation speed about twice that of rock. 

This means that pebbles should be roughly 4 times smaller inside the ice line than outside, in our model, compared to a factor of 10 found by \citet{morbidelli:2015}. Despite this difference in pebble size, the best-fit simulations in Section~4 yield essentially the same result as \citet{morbidelli:2015}: bodies inside and outside the ice line have masses of roughly 0.1 and 10 Earth masses, respectively. Since pebble accretion typically becomes more effective for larger pebbles, this implies that pebble accretion is inherently less efficient in our model, at least inside the ice line. This can be attributed to the inclusion of the gas advection term in Eqn.~\ref{eq-upeb} for the pebble radial velocity, which was not considered by \citet{morbidelli:2015}. Including gas advection substantially reduces the pebble surface density in the inner disk , lowering the pebble accretion rate in the same way that the presence of very small rocky pebbles would (see Figure~9).

\citet{morbidelli:2015} considered embryos with initial masses $\membo$, similar to the Moon. However, models and observations of small body populations suggest that planetesimals were smaller than this when they first formed---a few hundred km at most \citep{morbidelli:2009, simon:2016}. With this in mind, we used an initial embryo mass of $10^{-5}M_\oplus$, equivalent to a diameter of  350 km, while the other planetesimals are 100 km in diameter. 

For embryos this small, pebble accretion can be very slow because encounters with an embryo don't last long enough for gas drag to be effective. This is the reason for the exponential term in Eqn.~\ref{eq-tgrow}. Since the growth {\em rate} increases exponentially with mass, embryos that are somewhat larger than $\membo$ can grow very fast. This implies embryos will either grow to the pebble isolation mass or remain near $\membo$. Intermediate-mass bodies like Mars should be very rare. A similar argument was made by \citet{brasser:2020}.

However, the picture changes substantially when planetesimal accretion is included. Planetesimal accretion leads to steady growth interior to the ice line. Beyond the ice line, planetesimal accretion is slow, but it is faster than pebble accretion for small embryos.  Thus, planetesimal accretion can provide the boost needed for rapid pebble accretion to occur later in the evolution. As shown in Figure~12, planetesimal accretion allows multiple Mars-sized objects to form in the terrestrial-planet region, and it expands the region where giant-planet cores can form in the outer disk.

\citet{izidoro:2021b} have also examined the relative contributions of planetesimal and pebble accretion to the formation of the Sun's terrestrial planets. They looked at embryo growth at 0.5 and 1 AU in a disk with an imposed pressure bump at 5 AU that formed in the first 0.3 My of the disk's history. Such a bump could have arisen due to a large embryo or been a long-lived feature of the solar nebula. These authors found that planetesimal accretion dominated growth at 1 AU. This was also true at 0.5 AU for disks with strong turbulence. However, for weak turbulence, similar to the best-fit case considered here, \citet{izidoro:2021b} found rapid pebble accretion at 0.5 AU, forming Earth-mass planets in a few hundred ky. This difference with the current work can be attributed to the larger minimum pebble Stokes number needed for planetesimal formation used here, which delays the formation of embryos at 0.5 AU.

The planetesimal formation model used here was also used by \citet{lenz:2020} to calculate the initial planetesimal surface density $\sigmapml$ in the solar nebula. These authors compared $\sigmapml$ with the mass needed to form the observed planets and small-body populations at 7 locations in the Solar System. This allowed the authors to constrain uncertain parameters such as the planetesimal formation efficiency $\epsilon$, the initial disk mass $\mdisko$, and the fragmentation velocity $\vfrag$, which was assumed to be the same for ice and rock. 

\citet{lenz:2020} found that a wide range of parameter values produced a good match to the Solar System. Like the simulations described here, they also found that planet formation close to the Sun was slowed or prevented due to small values of the pebble Stokes number. The best-fit case found by \citet{lenz:2020} has $\epsilon=0.05$, $\mdisko$ and $\vfrag=200$ cm/s. These are similar to the values found in this study if we take $\vice$ for the fragmentation velocity. Their preferred value of the turbulence strength $\alphaturb=3\times 10^{-4}$ is larger than in this study, but the authors noted that a reasonable fit can be obtained for values as small as $\alphaturb=10^{-5}$. The minimum Stokes number $\stmin$ required for planetesimal formation was also similar to the one found here, allowing for the fact that \citet{lenz:2020} used a smooth decrease in formation efficiency rather than an abrupt transition.

These similarities are encouraging. However, we note that the current study provides much stronger constraints on the formation of the Solar System. Although \citet{lenz:2020} considered the mass of planetesimals needed to form the planets, they did not allow for the time required to grow the planets by accreting the planetesimals. They also did not consider pebble accretion, which was probably the dominant growth mode in the outer disk. These factors were included in the current study, and this explains why we find that good fits  occupy a much narrower region of phase space (see Figure~1).

%
% PLANETARY COMPOSITIONS
%
\subsection{Planetary Compositions}
Recently, it has become apparent that meteorites, and by extension their parent bodies, can be divided into two distinct groups on the basis of their isotopic compositions \citep{kleine:2020}. These are typically referred to as the CC (carbonaceous chondrite) and NC (non-carbonaceous) groups. The two groups of parent bodies have a range of ages that overlap each other. This indicates the parent bodies or their precursors formed in two regions of the solar nebula that co-existed but did not exchange much material. \citet{kruijer:2017} have proposed that the formation of Jupiter's core generated a barrier between the inner and outer Solar System when it reached the pebble isolation mass. This formed two isolated reservoirs, one inside the barrier, and one outside, that gave rise to NC and CC meteorite groups respectively.

Conversely, \citet{brasser:2020} calculated that Jupiter's growth by pebble accretion was too slow to form an effective barrier at the time required. They estimate that Jupiter's core took roughly 1 My to reach the pebble isolation mass. During this time, enough pebbles drifted past proto-Jupiter to contribute roughly 50\% of the total bulk of the terrestrial planets. This appears to be at odds with models that suggest that CC material makes up no more than a few percent of Earth by mass \citep{marty:2012}. 

In the simulations in Section 4, pebbles only contribute 10--20\% of the mass of embryos inside the ice line at 3 My (see Figure 11). The rest of the embryo mass comes from planetesimal accretion which was not considered by \cite{brasser:2020}. This fraction falls further if we allow for the sweep up of residual planetesimals after the solar nebula has dispersed. In this case, the pebble contribution becomes $<5\%$ for embryos in the region 1--1.5 AU.

However, this debate misses an important point. If planetesimals form continually from inward drifting pebbles, it makes no sense to distinguish between the contributions from pebble and planetesimal accretion. For example, pebbles that were accreted at 0.4 My would have had an identical composition to planetesimals that formed at 0.4 My. If these planetesimals were also accreted by an embryo, their effect on the embryo's composition would be the same as the pebbles themselves.

A better explanation for the meteorite dichotomy is that the composition of inward drifting pebbles changed over time. The parent bodies of the NC meteorites, together with the terrestrial planets, sampled pebbles and planetesimals that formed early, before the barrier was established. The parent bodies of the CC meteorites sampled pebbles and planetesimals that formed over a wider range of times, possibly dominated by material that formed after the barrier formed.

\citet{brasser:2020} have pointed out that the evolution of the ice line plays an important role in determining the masses of the terrestrial planets because it controls whether pebbles in this region are rocky or icy. This in turn determines the fragmentation velocity and Stokes number of the pebbles, and the rate of pebble accretion.

We extend this conclusion by noting that the location of the ice line at the time of pebble isolation is the key factor in determining the characteristics of the inner planets. A similar prediction was made by \citet{morbidelli:2016}. In the absence of strong viscous heating, the ice line is likely to enter the terrestrial-planet region at an early stage. Thus, the low masses of the terrestrial planets and their nearly ice-free compositions both require that pebble isolation occurred early in the evolution of the solar nebula.
   
%
% MODEL LIMITATIONS
%
\subsection{Model Limitations}
In the previous sections we have examined the early stages of planetary growth in the Solar System using a simple model of planet formation. This avoided physical processes that remain highly uncertain, and allowed us the study a wide range of parameter space. However, the simplifications mean the results should be treated with caution. Here, we briefly note some modeling caveats.

The model uses a specified, time-varying flux of fully formed pebbles injected into the outer edge of the model grid at 30 AU. A detailed model of pebble growth is not included, and the pebble size at each location is assumed to be the maximum size that pebbles can attain before collisions cause fragmentation rather than sticking.

We chose not to use a detailed dust growth model because existing calculations tend to predict the rapid loss of solids from the outer parts of disks \citep{brauer:2007}, in disagreement with observations of dust-rich mature disks \citep{ansdell:2016}. Instead, we assume that a significant reservoir of solid material survives until late times, but the pebble flux declines over time in common with the gas flux. We note that these two fluxes do not necessarily have a fixed ratio as assumed here.

The assumption that pebbles grow rapidly with most of the mass concentrated near a maximum size is supported by detailed growth models \citep{brauer:2008}. The maximum size can be set by either fragmentation or the radial drift rate \citep{birnstiel:2012}. The latter typically applies in the outermost regions of a disk or when pebble Stokes numbers are very large. Fragmentation limited growth is more relevant for the cases considered here. 

In the model used here, the inward flux of pebbles is halted when a planetary embryo reaches the pebble isolation mass. In reality, some small particles may be able to cross the partial gap in the disk opened by such an embryo \citep{weber:2018}. Thus, pebble isolation is unlikely to completely shut off the inward flow of solid material as we have assumed.

The orbits of planets can migrate radially due to tidal interactions with disk material. However, planetary migration is sensitive to the thermodynamics and flow of material close to the planet. Various modeling efforts have arrived at different conclusions, so that the rate and direction of migration are uncertain in many cases \citep{paardekooper:2011, lega:2014, benitez-llambay:2015, guilera:2021}. For this reason we chose to follow some other studies such as \citet{morbidelli:2015} and neglect migration.

Orbital migration is likely to be important in at least some systems. For example, it is a plausible explanation for the short orbital periods of many super Earths \citep{izidoro:2021a}. It is less clear how important migration was in the Solar System. In the terrestrial plant region planetary embryos may have been too small to undergo much migration while the solar nebula was present. Migration may have been a significant factor for the giant planets, although studies suggest the inward migration of Jupiter was limited by the presence of Saturn \citep{masset:2001, morbidelli:2007}. We note that Jupiter's core could have formed somewhat further out than its current location and still had time to accrete gas within the model used here. A different embryo, located closer to the Sun, at say 2.5 AU, could have been responsible for shutting off the pebble supply to the inner disk and causing the compositional dichotomy seen in meteorites, without necessarily growing into a giant planet. However, the fate of such an embryo is unclear and not addressed by this study.

Finally, we note that gas accretion onto planetary embryos is not included in the model. Currently, there are large uncertainties in the gas accretion rate for massive planets due to 3D hydrodynamical effects that are still being studied \citep{szulagyi:2014,lambrechts:2019}. For this reason, we chose to consider only whether embryos grow large enough to initiate gas accretion rather than estimating their final masses. We focus on Jupiter alone since the presence of Jupiter may have strongly influenced core formation and gas accretion rates for planets that formed later and further from the Sun.

%
% SUMMARY
%
\section{Summary}
Here, we presented a simple model for the early stages of planet formation in the Solar System with the evolution of the solar nebula driven by a disk wind. The model includes planetesimal formation from pebbles, and the growth of planetary embryos by the accretion of both pebbles and planetesimals. 

The model contains 6 parameters with large uncertainties: (i) the strength of the disk wind, (ii) the turbulence strength, (iii) the optical depth of the disk layer heated by the wind, (iv) the fragmentation strength of icy pebbles, (v) the efficiency of planetesimal formation, and (vi) the minimum size (Stokes number) of pebbles that can form planetesimals.

The values of these parameters were varied in order to match 6 features of the Solar System: (i) the total mass of the terrestrial planets, (ii) the small mass of Mercury, (iii) the water content of the terrestrial planets, (iv) the mass of a solid core needed to form Jupiter by runaway gas accretion, (v) the estimated mass of the primordial Kuiper belt, and (vi) the early formation of a barrier separating solid material in the inner and outer solar nebula.

The main findings are
\begin{itemize}
\item The model is able to satify all 6 of the Solar System constraints for a narrow range of parameter values, given in table 1.

\item The final masses of planetary embryos beyond the ice line are dominated by pebble accretion. Inside the ice line, pebble accretion can make a significant contribution to planetary growth, although planetesimal accretion is more important.

\item Pebble accretion growth rates are slower inside the ice line than outside for several reasons: (i) less solid material is available due to evaporation of the icy component of pebbles; (ii) pebble Stokes numbers are smaller, which reduces the pebble capture radius of embryos and increases the thickness of the pebble layer; (iii) a significant fraction of inflowing pebbles are filtered out by pebble accretion and planetesimal formation before they reach the inner disk; (iv) pebble advection speeds are increased by gas advection, which reduces the pebble surface density. A combination of these factors outweighs faster growth due to the shorter orbital periods inside ice line.

\item As the disk cools over time, the ice line moves inwards. The flux of pebbles to the inner disk is terminated once an embryo reaches the pebble isolation mass. If pebble isolation happens before the ice line crosses into terrestrial-planet region then the inner planets are essentially ice free, as observed.

\item Pebble accretion rates are very slow for the initial embryo mass used here. Rates increase rapidly as embryos grow larger. In the absence of planetesimal accretion, we get a sharp dichotomy in the final embryos masses: these are either near the pebble isolation mass or near the initial mass. This dichotomy is greatly reduced by planetesimal accretion, which allows Mars-mass embryos to form inside the ice line. Planetesimal accretion also allows embryos beyond the ice line to grow enough for rapid pebble accretion to take over, forming giant-planet cores over an extended region of the disk.

\item In the model used here, embryos and planetesimals can only form if the pebble Stokes number $\st$ exceeds a minimum value $\stmin$. In the region occupied by Mercury, $\st$ is small due to high temperatures and vigorous turbulence. As a result, planets are prevented from forming in much of region inside 0.5 AU. The only embryos to form here do so at a late stage, when the disk has cooled, and their growth is stunted by the onset of pebble isolation. These objects remain small as a result.

\end{itemize}

I would like to thank an anonymous reviewer for helpful comments that improved this paper.


\begin{thebibliography}{}
\bibitem[Ali-Dib \& Thompson(2020)]{ali-dib:2020} Ali-Dib, M. \& Thompson, C.\ 2020, \apj. 900, 96
\bibitem[Ansdell et al.(2016)]{ansdell:2016} Ansdell, M. et al.\ 2016, \apj, 828, 46
\bibitem[Benitez-Llambay et al.(2015)]{benitez-llambay:2015} Benitez-Llambay, P., Masset, F., Koenigsberger, G. \& Szulagyi, J.\ 2015, Nature, 520, 63
\bibitem[Birnstiel et al.(2012)]{birnstiel:2012} Birnstiel, T., Klahr, H. \& Ercolano, B.\ 2012, \aap, 539, A148
\bibitem[Bitsch et al.(2018)]{bitsch:2018} Bitsch, B.\, Morbidelli, A.\, Johansen, A.\, Lega, E., Lambrechts, M. \& Crida, A.\ 2018, \aap, 612, A30
\bibitem[Blum \& Wurm(2008)]{blum:2008} Blum, J. \& Wurm, G.\ 2008, Ann. Rev. Astron. Astrophys., 46, 21
\bibitem[Brauer et al.(2007)]{brauer:2007} Brauer, F., Dullemond, C.P., Johansen, A., Henning, Th., Klahr, H. \& Natta, A.\ 2007, \aap, 469, 1169
\bibitem[Brauer et al.(2008)]{brauer:2008} Brauer, F., Dullemond, C.P. \& Henning, Th.\ 2008, \aap, 480, 859
\bibitem[Brasser \& Mojzsis(2020)]{brasser:2020} Brasser, R, \& Mojzsis, S.J.\ 2020, Nat. Astron., 4, 492. 
\bibitem[Chambers(2009)]{chambers:2009} Chambers, J.E.\ 2009, \apj, 705, 1206
\bibitem[Chambers(2014)]{chambers:2014} Chambers, J.E.\ 2014, Icarus, 233, 83
\bibitem[Chambers(2017)]{chambers:2017} Chambers, J.E.\ 2017, \apj, 849, 30
\bibitem[Chambers(2018)]{chambers:2018} Chambers, J.E.\ 2018, \apj, 865, 30
\bibitem[Chiang \& Goldreich(1997)]{chiang:1997} Chiang, E.I. \& Goldreich, P.\ 1997, \apj, 490, 368
\bibitem[Clement et al.(2018)]{clement:2018} Clememt, M.S., Kaib, N.A., Raymond, S.N. \& Walsh, K.J.\ 2018, Icarus, 311, 340
\bibitem[Dullemond et al.(2018)]{dullemond:2018} Dullemond, C.P., Birnstiel, T., Huang, J. et al.\ 2018, \apj, 869, L46
\bibitem[Guilera et al.(2021)]{guilera:2021} Guilera, O.M., Miller, B., Mearcelo, M., Masset, F., Cuadra, J., Venturini, J. \& Ronco, M.P.\ 2021, \mnras, 507, 3638
\bibitem[Guilllot et al.(2014)]{guillot:2014} Guillot, T., Ida, S. \& Ormel, C.W.\ 2014, \aap, 572, A72
\bibitem[Gundlach \& Blum(2015)]{gundlach:2015} Gundlach, B. \& Blum, J.\ 2015, \apj, 798, 34
\bibitem[Guttler et al.(2010)]{guttler:2010} Guttler, C., Blum, J., Zsom, A., Ormel, C.W. \& Dullemond, C.P.\ 2010, \aap, 513, A56
\bibitem[Halliday(2013)]{halliday:2013} Halliday, A.N.\ 2013, Geochim. Cosmochim. Acta, 105, 146
\bibitem[Hartmann \& Bae(2018)]{hartmann:2018} Hartmann, L. \& Bae, J.\ 2018, \mnras, 474, 88
%\bibitem[Hubickyj et al.(2005)]{hubickyj:2005} Hubickyj, O., Bodenheimer, P. \& Lissauer, J.J.\ 2005, Icarus, 179, 415
\bibitem[Ikoma et al.(2000)]{ikoma:2000} Ikoma, M., Nakazawa, K. \& Emroi, H.\ 2000, \apj, 537, 1013
\bibitem[Izidoro et al.(2021a)]{izidoro:2021a} Izidoro, A., Bitsch, B., Raymond, S.N., Johansen, A., Morbidelli, A., Lambrechts, M. \& Jacobsen, S.A.\ 2021a, \aap, 650, A152
\bibitem[Izidoro et al.(2021b)]{izidoro:2021b} Izidoro, A., Bitsch,B. \& Dasgupta, R.\ 2021b, \apj, 915, 62
\bibitem[Johansen et al.(2007)]{johansen:2007} Johansen, A., Oishi, J.S., Mac Low, M.-M., Klahr, H., Henning, T. \& Youdin, A.\ 2007, Nature,  448m 1022
\bibitem[Johansen et al.(2012)]{johansen:2012} Johansen, A., Youdin, A.N. \& Lithwick,Y.\ 2012, \aap, 537, A125
\bibitem[Johansen et al.(2015)]{johansen:2015} Johansen, A., Mac Low, M.-M., Lacerda, P. \& Bizzarro, M.\ 2015, Sci. Adv. 1, e1500109.
\bibitem[Johansen et al.(2021)]{johansen:2021} Johansen, A., Thomas, R., Bizzarro, M., Schiller, M., Lambrechts, M., Nordlund, A. \& Lammer, H.\ 2021, Sci Adv. 7, eabc0444
\bibitem[Lega et al.(2014)]{lega:2014} Lega, E., Crida, A., Bitsch, B. \& Morbidelli, A.\ 2014, \mnras, 440, 683
\bibitem[Kleine et al.(2020)]{kleine:2020} Kleine, T., Budde, G., Burjhardt, C. ,Kruijer, T.S., Worsham, E.A., Morbidelli, A. \& Nimmo, F.\ 2020, Space Sci. Rev., 216, 55
\bibitem[Kokubo \& Ida(1998)]{kokubo:1998} Kokubo, E. \& Ida, S.\ 1998, Icarus, 131, 171
\bibitem[Kondo et al.(2022)]{kondo:2022} Kondo, K., Satoshi, O. and Mori, S.\ 2022, eprint arXiv:2205.13511
\bibitem[Kruijer et al.(2017)]{kruijer:2017} Kruijer, T.S., Burkhardt, C., Budde, G. \& Kleine, T.\  2017, Proc. Nat. Acad. Sci, 114, 6712
\bibitem[Lambrechts et al.(2012)]{lambrechts:2012} Lambrechts, M., \& Johansen, A.\ 2012, \aap, 544, A32
\bibitem[Lambrechts et al.(2014)]{lambrechts:2014} Lambrechts, M., Johansen, A. \& Morbidelli, A.\ 2014, \aap, 572, A35
\bibitem[Lambrechts et al.(2019)]{lambrechts:2019} Lambrechts, M, Lega, E., Nelson, R.P., Crida, A. \& Morbidelli, A.\ 2019, \aap, 630, A82
%\bibitem[Lee \& Chiang(2015)]{lee:2015} Lee, E.J. \& Chiang, E.\ 2015, \apj, 811, 41
\bibitem[Lenz et al.(2019)]{lenz:2019} Lenz, C.T., Klahr, H. \& Birnstiel, T.\ 2019, \apj, 874, 36
\bibitem[Lenz et al.(2020)]{lenz:2020} Lenz, C.T., Klahr, H. \& Birnstiel, T., Kretke, K. \& Stammler, S.\ 2020, \aap, 640, A61
\bibitem[Li et al.(2019)]{li:2019} Li, R., Youdin, A.N. \& Simon, J.B.\ 2019, \apj, 885, 69
\bibitem[Marty(2012)]{marty:2012} Marty, B., 2012, Earth Plan. Sci. Lett. 313, 56
\bibitem[Masset \& Snellgrove(2001)]{masset:2001} Masset, F. \& Snellgrove, M.\ 2001, \mnras, 320, L55
\bibitem[Morbidelli \& Crida(2007)]{morbidelli:2007} Morbidelli, A. \& Crida, A.\ 2007, Icarus, 191, 158
\bibitem[Morbidelli et al.(2009)]{morbidelli:2009} Morbidelli, A., Bottke, W.F. Nesvorny, D. \& Levison, H.F>\ 2009, Icarus, 204, 558
\bibitem[Morbidelli et al.(2015)]{morbidelli:2015} Morbidelli, A., Lambrechts, M., Jacobson, S. \& Bitsch, B.\ 2015, \icarus, 258, 418
\bibitem[Morbidelli et al.(2016)]{morbidelli:2016} Morbidelli, A., Bitsch, B., Crida, A., Gounelle, M., Guillot, T., Jacobson, S., Johansen, A, Lambrechts, M. \& Lega, E. \ 2016, \icarus, 267, 368
\bibitem[Musiolik \& Wurm(2019)]{musiolik:2019} Musiolik, G. \& Wurm, G.\ 2019, \apj, 873, 58
\bibitem[Nesvorny(2018)]{nesvorny:2018} Nesvorny, D.\ 2018, Ann. Rev. Astron. Astrophys., 56, 137
\bibitem[Ormel \& Cuzzi(2007)]{ormel:2007} Ormel, C.W. \& Cuzzi, J.N.\ 2007, \aap, 466, 413
\bibitem[Ormel \& Klahr(2010)]{ormel:2010} Ormel, C.W. \& Klahr, H.H.\ 2010, \aap, 520, A43
\bibitem[Paardekooper et al.(2011)]{paardekooper:2011} Paardekooper S.J., Baruteau, C. \& Kley, W.\ 2011, \mnras, 410, 293
\bibitem[Pollack et al.(1996)]{pollack:1996} Pollack, J.B., Hubickyj, O., Bodenheimer, P., Lissauer, J.J., Podolak, M. \& Greenzweig, Y.\ 1996, Icarus, 124, 62
\bibitem[Raymond et al.(2014)]{raymond:2014} Raymond, S.N., Kokubo, E., Morbidelli, A., Morishima, R. \& Walsh, K.J.\ 2014, Protostars and Planets VI, Henrik Beuther et al. eds, Univ. Arizona Press, pp595
\bibitem[Simon et al.(2016)]{simon:2016} Simon, J.B., Armitage, P.J., Li, R. \& Youdin, A.N.\ 2016, \apj, 822, 55
\bibitem[Suzuki et al.(2016)]{suzuki:2016} Suzuki, T.K., Ogihara, M. Morbidell, A., Crida, A. \& Guillot, T.\ \aap, 596, A74
\bibitem[Szulagyi et al.(2014)]{szulagyi:2014} Szulagyi, J., Morbidelli, A., Crida, A. \& Masset, F.\ 2014, \apj, 782, 65
\bibitem[Tsiganis et al.(2005)]{tsiganis:2005} Tsiganis, K., Gomes, R., Morbidelli, A, \& Levison, H.F.\ 2005, Nature, 435, 459
\bibitem[Tychoniec et al.(2020)]{tychoniec:2020} Tychoniec, L. et al.\ 2020, \aap, 640, A19
\bibitem[Walsh et al.(2011)]{walsh:2011} Walsh, K.J., Morbidelli, A., Raymond, S.N., O'Brien, D.P. \& Mandell, A.M.\ 2011, Nature, 475, 206
\bibitem[Ward(1997)]{ward:1997} Ward, W.R.\ 1997, Icarus, 126, 261
\bibitem[Weber et al.(2018)]{weber:2018} Weber, P., Benitez-Llambay, P., Gressel, O. Krapp, L. \& Pessah, M.E.\ 2018, \apj, 854, 153
\bibitem[Weidenschilling(1977)]{weidenschilling:1977} Weidenschilling, S.J.\ 1977, \mnras, 180, 57
\bibitem[Youdin \& Lithwick(2007)]{youdin:2007} Youdin, A.N. \&.Lithwick, Y.\ 2007, Icarus, 192, 588
\end{thebibliography}
\end{document}